\documentclass[aps,pre,preprint,amsmath,amssymb,nofootinbib]{revtex4-2}
\usepackage[utf8]{inputenc}
\providecommand{\DeclareUnicodeCharacter}[2]{}
\DeclareUnicodeCharacter{00B5}{\ensuremath{\mu}}
\DeclareUnicodeCharacter{2212}{\ensuremath{-}}
\DeclareUnicodeCharacter{2020}{\textdagger}
\DeclareUnicodeCharacter{00D7}{\ensuremath{\times}}
\DeclareUnicodeCharacter{2014}{---}
\DeclareUnicodeCharacter{00ED}{\'{\i}}
\DeclareUnicodeCharacter{00E1}{\'a}
\DeclareUnicodeCharacter{00F3}{\'o}
\DeclareUnicodeCharacter{00E0}{\`a}
\DeclareUnicodeCharacter{00F8}{\o}
\DeclareUnicodeCharacter{0142}{\l}
\DeclareUnicodeCharacter{0141}{\L}
\usepackage{graphicx}
\usepackage[export]{adjustbox}
\usepackage{bm}
\usepackage{longtable,booktabs,array}
\usepackage{calc}
\usepackage[colorlinks=true,citecolor=blue,linkcolor=blue,urlcolor=blue]{hyperref}

\makeatletter\let\auto@bib@innerbib\@empty\makeatother
\newif\iftwocolumnbuild
\twocolumnbuildfalse 
\begin{document}
\title{Reproducible Instance-Dependent Failure in Quantum Annealing: Predictability and Configuration-Space Geometry}
\author{Natalia Romero}
\affiliation{Machine Perception and Cognitive Robotics Laboratory, Florida Atlantic University\\Natural and Artificial Intelligence Institute}
\begin{abstract}
Quantum annealers realize a programmed Ising Hamiltonian only approximately, and their probability of reaching the ground state varies strongly between instances drawn from the same ensemble. Using structured, embedded random-field Ising instances with exactly certified ground states (3,158 instances, up to 360 spins), we ask whether this instance dependence is reproducible and predictable, pre-registering 87 predictions across two cells of a field-to-coupling parameter space, of which 61 held (45 of 71 once near-certain-by-design stability checks are excluded). Occupation varies nearly across its full range at fixed ensemble parameters, falls below simulated annealing in most instances, and both the hardness ranking and its collapse with system size reproduce on a second processor. Two pre-registered physical features predict failure before hardware contact, each confirmed at only one of two parameter regimes: ground-state magnetization at a calibrated point, and net field at a low-field corner, which loses significance once embedding quality is controlled for and fails to replicate in a dedicated, confound-controlled follow-up. A hardware-observed competing state's distance from the ground state, and an energy barrier estimated from it, are statistically associated with an instance's own occupation beyond these known features, replicating on held-out and freshly collected instances across two further processors. A classical model of small, documented-scale control-error offsets explains part of this predictability at the low-field corner without identifying the error map actually realized on the hardware. Predictability is real but weak and regime-dependent; the more robust finding is geometry's replicated, confound-controlled link to occupation, with value in design lessons rather than operational triage.
\end{abstract}
\maketitle
\section{Introduction}\label{introduction}

A quantum annealer is a programmable physical system: it realizes a user-specified Ising Hamiltonian on superconducting flux qubits and returns low-energy configurations of it. How faithfully it does
so is a question about the device as much as about the algorithm. D-Wave's own documentation reports that the physically realized fields and couplings differ from the programmed values by integrated
control errors (ICE), typically about 1\% and up to 2--4\% of the programmable range, arising from background susceptibility, flux noise, digital-to-analog quantization, I/O effects and crosstalk
\cite{ref1}. On native spin-glass benchmarks, Zhu \emph{et al.} \cite{ref5} showed that instances with identical noise statistics can have very different success probabilities, a chaotic sensitivity
of the ground state to small coupling perturbations that the earlier spin-glass benchmarking literature had already termed J-chaos; a related phenomenon, heavy-tailed hardness distributions dominated
by a small subset of instances, has been traced to high local degeneracy and late-anneal perturbative crossings \cite{ref49}. Pearson \emph{et al.} \cite{ref2} later showed, by adding Gaussian
perturbations at the ICE scale to \(h\) and \(J\), that such errors make an annealer return ``the correct answer to the wrong problem.'' Comparisons of annealers with simulated annealing (SA)
date to the first hundred-qubit devices \cite{ref3}, where hard instances were linked to small-gap avoided level crossings -- a mechanism for which specific hard-instance constructions can make the
gap decay superexponentially with system size \cite{ref45} -- and hardware success tracked simulated quantum annealing rather than SA; the
same and follow-up work found that, among runs that miss the ground state, the competing state sits a small Hamming distance away for easy instances and a large one for hard instances
\cite{ref3,ref42} --- the landscape signature we reproduce below (Sec. III E--F) on embedded, structured instances with exactly certified ground states, rather than the native, unembedded benchmarks
used there. Martín-Mayor and Hen \cite{ref4} showed that classically hard instances are those whose low-lying excitations lie far from the ground state, and Mehta \emph{et al.} \cite{ref40} found
that the scaling exponents of annealer success probability on small QUBO instances correlate with the Hamming-distance distribution of their low-lying states --- the closest prior result to our own
landscape-geometry finding (Sec. III F), though at the level of ensemble scaling exponents rather than per-instance prediction conditioned on classical hardness and a physical predictor. Stated
plainly: the landscape-hardness link itself is not new, and we do not claim otherwise. Boixo \emph{et al.} established it on native, unembedded benchmarks as a qualitative signature; Martín-Mayor and
Hen established it as a classical-hardness criterion; Mehta \emph{et al.} established it at the level of ensemble scaling exponents. The inference this work adds is a per-instance, conditional
association -- controlling for classical hardness and each cell's preregistered physical predictor, not a bivariate one -- replicated on structured, embedded instances with exactly certified ground states and
validated prospectively on freshly collected, never-selected-on-outcome data (Sec. III F). Koshka and
Novotny \cite{ref6} found that annealer and SA sample different local minima of the same landscape, so the two
are not merely noisier and cleaner estimators of one distribution. Finite-temperature freeze-out of the transverse-field dynamics is a further, independent source of deviation from ground-state
sampling \cite{ref7}.

These results leave open a question that matters for any use of an annealer as a sampler: given an ensemble of instances, can one predict \emph{which} instances will fail? Pearson \emph{et al.} left
this open, and most subsequent tests use synthetic benchmark ensembles that are native to the hardware graph. Certo \emph{et al.} \cite{ref8} predict annealer-versus-SA suitability from coefficient
distributions on such unembedded instances, Barbosa \emph{et al.} \cite{ref9} predict optimality on one problem class from simple instance features, and Pellini and Ferrari Dacrema \cite{ref41} train
several meta-learning classifiers on over a hundred instance-level features across five thousand instances of ten problem classes, finding that the distribution of the bias and coupling coefficients,
not their density alone, is informative of success; none of the three tests physical per-instance
properties on held-out hardware data; Sec. III C gives a direct, quantitative comparison between our physical predictors and a cheap-feature machine-learning baseline in their style. Positive scaling results concern native problems or approximate optimization \cite{ref10,ref11}, and runtime comparisons that count all overheads show no
advantage on current hardware \cite{ref12}.

Most prior studies of instance-dependent failure use synthetic spin glasses native to the hardware graph, with no embedding; in practice, users run structured problems that need embedding. Here we
study instance-dependent failure on structured, embedded instances that were not designed to probe hardware error: a random-field Ising spin glass on degree-capped Barabási--Albert graphs,
with fields calibrated to survey data and a controlled fraction of antiferromagnetic bonds. These instances have heterogeneous fields calibrated to real data, and let field-to-coupling ratio, density
and frustration be tuned separately, which is what revealed that the working predictor depends on the regime. The model belongs to a family of Ising-type descriptions of social and political systems, in which local fields represent individual predispositions and signed couplings represent alliances, rivalries or contrarian behavior, widely used both by Galam \cite{ref13,ref14,ref15} and in the broader statistical-physics literature on opinion dynamics \cite{ref29,ref30,ref31}; here it serves only as a source of non-synthetic instance ensembles with tunable field-to-coupling ratio, density and frustration, not as a model of opinion formation in its own right. Companion work characterizes this model's statistical mechanics in the social-network setting \cite{ref16,ref17}. We use the annealer throughout as a physical sampler of a fixed
Hamiltonian, not as a solver for an agent-based process, which avoids the encoding pitfalls identified by Barati \emph{et al.} \cite{ref18}.

This work makes six contributions. The ground state of 3,158 instances (\(N\le360\)) is certified globally optimal by exact optimization, so the outcome variable, ground-state occupation, is measured
against the true minimum rather than against a heuristic's best (Sec. II C). At fixed ensemble parameters, annealer occupation varies across nearly its full range, and a sweep of size, field-to-coupling ratio and frustration
shows where, and how fast, the annealer falls behind SA (Secs. III A--B). A held-out, pre-registered test finds that physical instance features predict failure, but which feature does so depends on
the parameter regime (Sec. III C). A classical static-error model reproduces part of the hardware's instance ranking on the same held-out sample (Sec. III D), and hardware diagnostics constrain the
alternatives (Sec. III E). A pre-registered robustness campaign of 87 held-out predictions tests whether these findings are properties of one chip and one moment in time -- across four days, three
processors, a new graph family and a new density, spin-reversal gauges, a second size-scaling run, and a dedicated, confound-controlled replication of the weaker predictor -- of which 61 held (45 of 71
once near-certain-by-design stability checks are excluded; Sec. III F). And the competing state's Hamming distance and an estimated
energy barrier predict an instance's own hardware occupation beyond classical hardness and the pre-registered predictor, confirmed on samples never selected on the outcome and on two processors
(Sec. III F).

The remainder of the paper is organized as follows. Section II defines the Hamiltonian, instance ensembles, hardware and embedding, and the exact reference ground states against which occupation is
measured. Section III presents the results: instance dependence at fixed parameters and across system size, field-to-coupling ratio and frustration (Secs. III A--B); the pre-registered predictor test
(Sec. III C); the static control-error model (Sec. III D); hardware diagnostics of the competing states (Sec. III E); the robustness campaign across days, processors, gauges and system size, including
the landscape-geometry predictor (Sec. III F); and the dynamical mechanism tests (Sec. III G). Section IV discusses the findings and their limitations, and Section V gives methods
and data availability.

\section{Model, hardware and reference ground states}\label{model-hardware-and-reference-ground-states}

\subsection{Hamiltonian and instance ensembles}\label{hamiltonian-and-instance-ensembles}

We study \[H(\mathbf{s}) = -\sum_i h_i s_i - \sum_{(i,j)} J_{ij} s_i s_j, \qquad s_i=\pm1,\] on degree-capped Barabási--Albert graphs \cite{ref19} with attachment parameter \(m\in\{2,5,10\}\) and
maximum degree 20, chosen to match the hardware's coupler limit. Degree capping lowers the realized mean degree below the nominal \(2m\) (12.4 rather than 20 at \(m=10\)). Fields \(h_i\) take seven
tiers from \(\pm1.729\) to 0, calibrated to partisan-strength responses in the ANES 2020 survey \cite{ref20}, with a 1.3\(\times\) multiplier on a random 10\% of sites. Coupling magnitudes take four
tiers (0.55, 0.40, 0.25, 0.10), ordered by tie strength; this ordering is stylized rather than measured \cite{ref21}. A fraction \(\alpha\) of bonds is antiferromagnetic, and \(r_h\) is a global
multiplier on the field scale (\(r_h=1\) is the calibrated scale). We call each fixed point in \((m,r_h,\alpha)\) parameter space a \emph{cell}, labeled by a letter (Table I); the parameter cells at
which the annealer is compared with SA are listed there. The two cells used for the pre-registered test
are the calibrated point E \(=(m,r_h,\alpha)=(10,1,0.20)\) and the low-field corner C \(=(10,0.25,0.20)\). An instance is \emph{hard} if its annealer ground-state occupation (defined in Sec. II C) is
low and \emph{easy} if it is high; the operational threshold differs by analysis (a fixed occupation cutoff in Sec. III C, the extremes of a held-out sample in Secs. III E--F) and is stated at first
use in each case.

\subsection{Hardware and embedding}\label{hardware-and-embedding}

Unless stated otherwise, submissions used \texttt{Advantage2\_system1} (Zephyr topology \cite{ref22}; 4,575 active qubits, 41,474 active couplers, queried live), with \(h\)-range \([-6,6]\), \(J\)-range \([-1,1]\), and a
default anneal time of 20 µs (Sec. III F additionally uses two Pegasus-topology processors, \texttt{Advantage\_system4} and \texttt{Advantage\_system6}). Instances were minor-embedded with \texttt{minorminer} against the live hardware graph and sampled through \texttt{FixedEmbeddingComposite}, so that embeddings could be
held fixed while schedules or gauges varied, or replaced while the instance was held fixed. The solver's sign convention was converted (\(h'=-h\), \(J'=-J\)) and checked by a standing test before
every reported run. Chain strength used the \texttt{uniform\_torque\_compensation} default unless swept.

\subsection{Exact reference ground states and the outcome variable}\label{exact-reference-ground-states-and-the-outcome-variable}

The outcome throughout is \emph{ground-state occupation}: the fraction of reads whose energy equals the instance's ground-state energy to within \(10^{-6}\) (absolute),
\[
Y = \frac{1}{R}\sum_{r=1}^{R} [\![\,|H(\mathbf{s}_r) - E_0| < 10^{-6}\,]\!],
\]
where \(\mathbf{s}_r\) is the configuration returned by read \(r\) of \(R\) total reads, \(E_0\) is the certified ground-state energy, and \([\![\cdot]\!]\) equals 1 if the enclosed condition holds and 0 otherwise. Reference energies were first
obtained by SA before any annealer submission (5,000 reads, 1,500 sweeps; \texttt{dwave-neal}), and were never revised using annealer output. We then computed every instance's ground state exactly, as
a binary quadratic program (substituting \(s_i = 2x_i-1\)) solved to zero optimality gap: with Gurobi 12.0.3 for \(N\le180\), and, for the 60 instances at \(N=360\), which exceed the size limit of the
license used, with the CP-SAT solver of OR-Tools 9.15 on couplings scaled by \(10^9\) and rounded. Rounding each of the \(N\) fields and \(|E|\) couplings to the nearest integer before scaling back changes
any configuration's energy by at most \(\Delta = 0.5/10^9 \times (N+|E|)\); for the 60 instances at \(N=360\), \(\Delta\) ranges \(4.86\times10^{-7}\)--\(5.21\times10^{-7}\) (recorded per instance), so the rounded
optimum equals the true optimum unless two configurations' energies differ by less than \(2\Delta\). We checked this directly rather than leaving it a possibility: re-solving each of the 60 instances with the
objective constrained strictly above the found optimum gives the exact gap, in the same rounded units, to the next-best distinct configuration; the smallest such gap across all 60 instances is
\(2.3\times10^{-4}\), at least 225 times \(2\Delta\) in every instance, so no rounding-induced perturbation of this size could have changed which configuration is optimal. \(\Delta\) is itself about half the \(10^{-6}\)
energy-matching tolerance used throughout for \(Y\) above. The two solvers agree to full
precision on cross-checked instances. All 3,158 instances (400 in the pre-registered sample, 1,220 in the parameter sweeps, 36 + 2 in the six-seed campaign and the two most-studied single instances,
100 in the sparse-graph check of Sec. III E, 1,200 in Sec. III F's three further cells, and the 200 fresh instances of Sec. III F's prospective landscape-geometry validation) returned a proof of
optimality within 50 s. The 300 additional Corner instances of the net-field replication (Sec. III C) are not included in this count: they use the same SA-reference convention as every other
campaign in this study, without separate exact certification.

The SA reference equals the certified ground-state energy in 3,131 of 3,158 instances, including all instances with \(N\le90\) and all instances at \(r_h=1\). The 27 exceptions all lie in the
weak-field, sparse corner (\(m=2\), \(r_h=0.25\), \(\alpha=0.20\)) at large size: 2 of 30 instances at \(N=180\) and 25 of 30 at \(N=360\), where SA's best state lies 0.001--0.17 above the ground
state. No annealer read reached the true ground state on any of these 27 instances either. We report occupation against the certified energy throughout, so both solvers' occupation is zero on these
instances. The correction changes no reported conclusion, because the affected occupations were already below 1\%. It does mean that at \(N=360\), \(r_h=0.25\) neither solver samples the ground state
at all within its budget. ``Ground state'' below means the certified global minimum. The \(10^{-6}\) equality tolerance is set against the measured level spacing: among the 400 held-out instances (\(N=90\); a separate check
from the \(N=360\) rounding margin above), the
smallest gap to the first excited level is \(3.4\times10^{-5}\) (one instance below \(10^{-3}\); the median gap is 0.09--0.11), so the tolerance is more than 30 times smaller than any observed level
spacing and cannot merge distinct energy levels. Degenerate ground states (14 instances at Empirical, at most 4 configurations; 8 at Corner, at most 2) are counted correctly under this tolerance, since every
configuration at the certified energy is a true co-optimum, not a near-miss.

\section{Results}\label{results}

\subsection{Ground-state occupation is strongly instance dependent at fixed parameters}\label{ground-state-occupation-is-strongly-instance-dependent-at-fixed-parameters}

Figure 1 shows annealer versus SA occupation for the 400 held-out instances of the pre-registered sample (200 each at the Corner and Empirical cells; four embeddings and 1,000 reads per embedding per instance). At
fixed ensemble parameters, annealer occupation spans nearly its full range: at the Corner cell it ranges from 0\% to 90\% (median 9.8\%); at the calibrated Empirical cell, 153 of 200 instances have occupation below
5\% and 18 have none. The annealer exceeds SA in 8 of 200 instances at Corner and none at Empirical (mean paired differences \(-26.0\) and \(-43.3\) percentage points). As a sanity check on the exact certification, since no valid read can fall below the true minimum, no annealer read in any of the 400 instances did so.

\begin{figure}[tbp]
\centering
\includegraphics[max width=\linewidth]{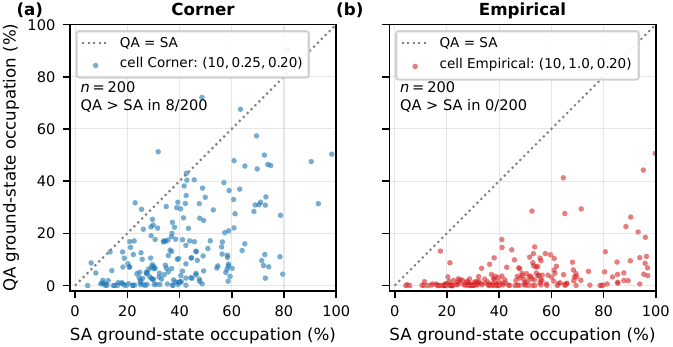}
\caption{Annealer (QA) versus simulated-annealing (SA) ground-state occupation for the 400 pre-registered, held-out instances at (a) the Corner cell and (b) the Empirical cell. Each point is one instance; QA occupation is the mean
over four embeddings of 1,000 reads. Dotted line, QA = SA. QA occupation is below SA occupation in 392 of the 400 instances.}
\end{figure}

The same spread is visible within a single parameter cell at small sample size. At the Corner cell, six realizations studied in detail gave occupations of 88.9\%, 74.9\%, 24.0\%, 4.2\%, 0.0\% and 0.0\% (one
embedding each); re-embedding an easy and a hard realization eight times each confirmed that the classification is a property of the instance, not of the embedding drawn (38.2\% vs.~0.42\% mean
occupation; SM S5). The easy realizations' dominant competing states lay 1--2 spins from the ground state; the three hardest lay 7, 83 and 86 spins (of 90) away.

Table I collects every annealer-versus-SA comparison in this study, labeled by campaign: ``held-out'' rows are the pre-registered, confirmatory 400-instance sample (Fig. 1); ``sweep'' rows are the
pre-specified axis tests of Sec. III B; ``six-seed'' rows are the exploratory pilot that motivated the held-out design and are not confirmatory evidence on their own. Small campaigns can give the
wrong sign. At the Corner cell, a six-realization pass showed the annealer ahead at 3 of 6 seeds (mean \(+2.9\)
pp), but the 200-instance sample reverses this. At \((10,0.25,0.05)\), two realizations studied first gave annealer occupations of 94.7\% and 99.2\% against SA's 55.4\% and 69.3\% (SM S2), but 20
fresh instances give parity. The two remaining six-seed-only favorable cells (rows marked †) were not retested at the same \(\alpha\). At \(\alpha=0.20\) the nearest swept cells give parity (\(m=2\))
or an annealer deficit (\(m=5\)).

\textbf{Table I.} Annealer (QA) vs.~SA ground-state occupation across campaigns, \(N=90\) unless stated. \(n\) = instances. Paired difference = QA \(-\) SA, mean (95\% bootstrap CI where available),
in percentage points. Holm correction is within each sweep. The table is grouped by evidentiary status: confirmatory (the pre-registered 200-instance held-out test), then exploratory campaigns
(parameter sweeps, \(n=20\)--30 per cell), then the six-seed pilot (\(n=6\) per cell, the smallest and least reliable rows, reported for the small-sample-reversal point of Sec. IV only).

{\footnotesize\setlength{\tabcolsep}{4pt}
\resizebox{\linewidth}{!}{%
\begin{tabular}{@{}llrrrrlr@{}}
\toprule
Campaign & Cell \((m, r_h, \alpha)\) & \(n\) & QA mean & SA mean & QA \textgreater{} SA & Paired diff. & Holm \(p\) \\
\midrule
Held-out (A1), confirmatory & Corner (10, 0.25, 0.20) & 200 & 14.8\% & 40.8\% & 8/200 & \(-26.0\) & --- \\
Held-out (A1), confirmatory & Empirical (10, 1.0, 0.20) & 200 & 4.3\% & 47.7\% & 0/200 & \(-43.3\) & --- \\
\midrule
\(\alpha\) sweep (S3), exploratory & (10, 0.25, 0.05) & 20 & 60.0\% & 57.8\% & --- & \(+2.2\) {[}\(-10.5\), \(+13.4\){]} & 0.55 \\
\(\alpha\) sweep (S3), exploratory & (10, 1.0, 0.05) & 20 & 10.0\% & 60.9\% & --- & \(-50.9\) {[}\(-58.8\), \(-42.7\){]} & \(2\times10^{-5}\) \\
Size sweep (S1), exploratory & (2, 0.25, 0.20) & 30 & 12.3\% & 11.1\% & --- & \(+1.2\) {[}\(-3.2\), \(+6.0\){]} & 1 \\
Size sweep (S1), exploratory & (5, 0.25, 0.20) & 30 & 12.1\% & 30.5\% & --- & \(-18.4\) {[}\(-24.5\), \(-13.1\){]} & \(1\times10^{-7}\) \\
\midrule
Six-seed pilot† & (2, 0.25, 0.05) & 6 & 34.8\% & 13.2\% & 4/6 & \(+21.6\) & --- \\
Six-seed pilot† & (5, 0.25, 0.05) & 6 & 62.9\% & 51.7\% & 4/6 & \(+11.2\) & --- \\
Six-seed pilot & (10, 0.25, 0.20) = Corner & 6 & 32.0\% & 29.1\% & 3/6 & \(+2.9\) & --- \\
Six-seed pilot & (10, 1.0, 0.05) & 6 & 11.2\% & 53.5\% & 1/6 & \(-42.3\) & --- \\
Six-seed pilot & Empirical & 6 & 15.3\% & 71.2\% & 0/6 & \(-55.9\) & --- \\
Six-seed pilot & (10, 2.0, 0.05) & 6 & 4.5\% & 62.3\% & 0/6 & \(-57.9\) & --- \\
\bottomrule
\end{tabular}}
}

\subsection{Dependence on system size, field-to-coupling ratio and frustration}\label{dependence-on-system-size-field-to-coupling-ratio-and-frustration}

A sweep campaign (S1--S3; about 7,000 submissions, four embeddings per instance, paired against SA on the same instances) maps where the annealer falls behind (Figs. 2--3). Across every swept cell at
\(N\ge90\), the annealer's mean occupation is at or below SA's. No cell shows a Holm-significant annealer advantage at any \(N\). At \(N=30\) the nominal largest advantage, \(+9.0\) pp at
\((2,0.25,0.20)\), does not survive correction (\(p_\text{Holm}=0.20\)).

\textbf{System size.} Both solvers' occupations fall with \(N\), the annealer's faster. Because \(N=360\) was run only at \(m=2\) (at \(m=10\) no embedding was found, and at \(m=5\) chains reach 28
qubits), we compare like with like at \(m=2\) (pooled over \(r_h\in\{0.25,1\}\), 30 instances per point): the annealer-to-SA occupation ratio is 1.04 at \(N=30\), 0.95 at \(N=60\), 0.78 at \(N=90\), 0.42 at \(N=180\)
and 0.07 at \(N=360\). Pooled over all densities the ratio is 0.82, 0.55, 0.36 and 0.09 at \(N=30\), 60, 90 and 180 (no other density exists at \(N=360\), so this all-densities view does not extend there).
Restricted to the one density available at both ends of the range, \(m=2\) (the same comparison as above, now tested at the instance level): unpaired Mann--Whitney, \(N=30\) vs.\ \(N=360\),
\(p=8.0\times10^{-16}\). This is a statement
about ground-state \emph{sampling probability} at a fixed 20 µs anneal and default settings, not a time-to-solution or computational-complexity comparison. Because the ratio compresses as both
occupations shrink, the absolute (pooled, \(m=2\)) means behind it are annealer/SA 59.2\%/56.8\% at \(N=30\), 32.1\%/33.9\% at \(N=60\), 20.4\%/26.2\% at \(N=90\), 3.7\%/8.7\% at \(N=180\) and 0.12\%/1.7\% at \(N=360\); at
\(N=360\), \(r_h=0.25\) both are zero (Sec. II C), so the \(N=360\) ratio rests on the \(r_h=1\) instances. On a second processor (Pegasus topology, same instances and references), this ratio collapse reproduces (Spearman \(\rho=-0.88\), \(p<10^{-4}\), pooled over \(r_h\) at \(m=2\)), and the per-instance ranking transfers at matched size, density and field ratio (positive, significant correlation at 19 of 19 testable parameter points out of 20; the remaining point is degenerate because both solvers return zero occupation on every instance there), extending the size-scaling finding beyond the processor used for Figs. 2--3.

\begin{figure}[tbp]
\centering
\includegraphics[max width=\linewidth]{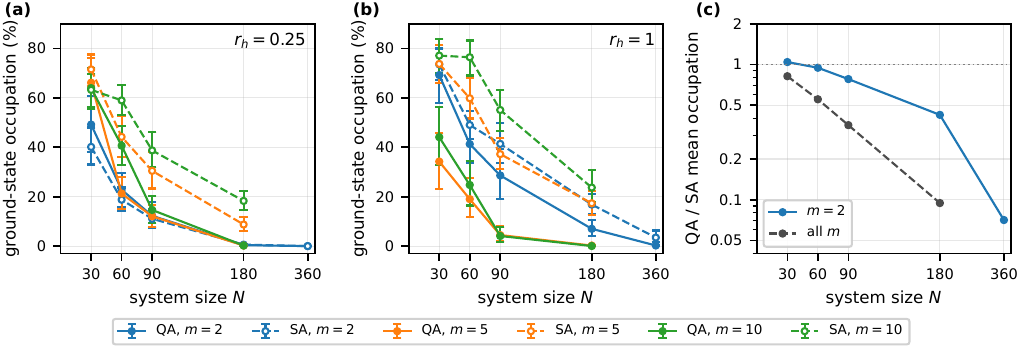}
\caption{Ground-state occupation versus system size \(N\) at \(\alpha=0.20\) for (a) \(r_h=0.25\) and (b) \(r_h=1\), at densities \(m=2,5,10\). Filled symbols and solid lines, QA; open symbols and dashed
lines, SA; error bars, bootstrap 95\% confidence intervals; 30 instances per point. (c) Ratio of mean QA to mean SA occupation at \(m=2\) (pooled over \(r_h\)) and pooled over all densities
(\(N\le180\)). Occupations are measured against certified ground states.}
\end{figure}

\textbf{Field-to-coupling ratio.} At \(N=90\) the deficit is largest at intermediate \(r_h\) and narrows at both extremes of the tested range \(r_h\in[0.1,4]\) (axis test \(p=3.5\times10^{-3}\); Fig.
3). At \(m=2\) the annealer reaches parity for \(r_h\le0.5\), while at \(m=10\) the deficit exceeds 45 pp for every \(r_h\ge1\).

\begin{figure}[tbp]
\centering
\includegraphics[max width=\linewidth]{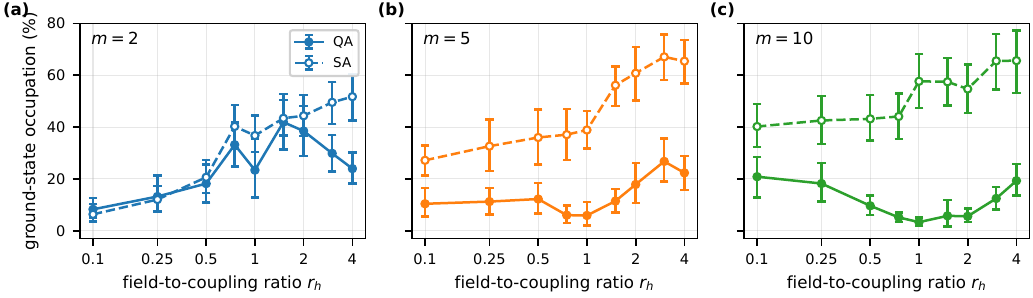}
\caption{Ground-state occupation versus field-to-coupling ratio \(r_h\) at \(N=90\), \(\alpha=0.20\), for (a) \(m=2\), (b) \(m=5\) and (c) \(m=10\); 20 instances per point. Symbols as in Fig. 2.}
\end{figure}

\textbf{Antiferromagnetic fraction.} Increasing \(\alpha\) from 0.05 to 1 lowers annealer occupation more steeply than any other swept parameter (axis test \(p=1.9\times10^{-6}\); SM Fig. S12). At
\(r_h=0.25\) the annealer falls from parity at \(\alpha=0.05\) to 0.8--2.1\% occupation for \(\alpha\ge0.35\), while SA retains 23--36\%. This collapse is not a universal property of \(\alpha\),
however: repeating the sweep at \(m=2\) and \(m=5\) (same \(N=90\), 20 instances per point) shows it is conditional on both density and field ratio. At \(r_h=0.25\) the paired drop from
\(\alpha=0.05\) to \(\alpha=1\) strengthens with density -- \(-26\) pp at \(m=2\) (\(p=2.3\times10^{-3}\)), \(-61\) pp at \(m=5\) (\(p=1.9\times10^{-6}\)), \(-59\) pp at \(m=10\)
(\(p=1.9\times10^{-6}\)) -- while at \(r_h=1\) the same comparison is far weaker and, at the lowest density, absent: \(-6\) pp at \(m=10\) (\(p=3.7\times10^{-3}\)), \(+1.5\) pp at \(m=5\)
(\(p=2.4\times10^{-2}\), a small effect in the opposite direction), and \(+4.9\) pp at \(m=2\) (\(p=0.50\), not significant). The parameter that drives hardness most strongly in this ensemble is
therefore not a universal lever: at low density and high field-to-coupling ratio it leaves occupation unchanged. Figure 4 shows this phase map directly, across density and field ratio.

\begin{figure}[tbp]
\centering
\includegraphics[max width=\linewidth]{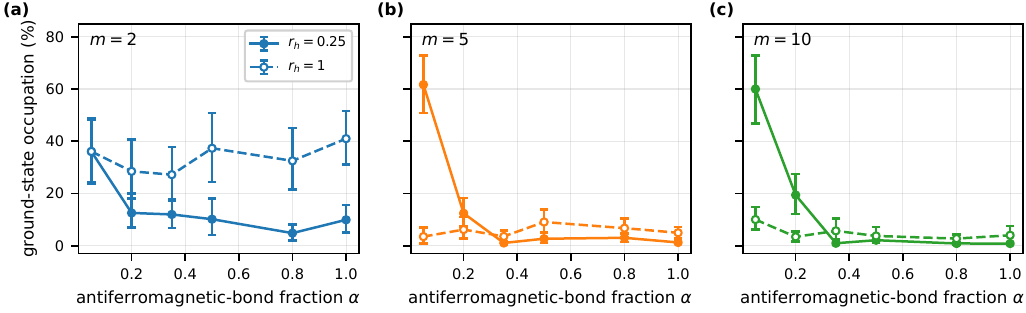}
\caption{Ground-state occupation versus antiferromagnetic-bond fraction \(\alpha\) at \(N=90\), for (a) \(m=2\), (b) \(m=5\) and (c) \(m=10\), each at \(r_h=0.25\) (solid) and \(r_h=1\) (dashed); 20
instances per point, error bars are bootstrap 95\% confidence intervals. The \(m=10\) panel reproduces SM Fig. S12; the \(m=2\) and \(m=5\) panels are new data from the same protocol. The steep
\(\alpha\)-driven collapse at \(r_h=0.25\) strengthens with density and nearly vanishes at \(r_h=1\), most completely at \(m=2\) (Sec. III B, ``Antiferromagnetic fraction'').}
\end{figure}

\textbf{Hardware controls.} Three control parameters shift occupation without closing the gap (SM S11): weaker chains help (0.5\(\times\) vs.~3\(\times\) default, \(p=2.0\times10^{-4}\)); using the
full energy scale beats every reduced scale (1.0 vs.~0.1, \(+4.5\) pp, \(p=1.9\times10^{-9}\)); and across anneal times of 0.5--2,000 µs no individual cell differs significantly, although the pooled
endpoint test (500 vs.~0.5 µs) finds a small effect (\(+4.8\) pp, \(p=2.4\times10^{-4}\)).

\subsection{Pre-registered, held-out predictors of failure}\label{pre-registered-held-out-predictors-of-failure}

An exploratory analysis of an earlier 240-instance data set (two cells, 120 instances each) found that ground-state magnetization \(|M|\) correlated with annealer occupation more strongly than with SA
occupation. Before collecting new data, we pre-registered four one-sided Spearman tests, of \(|M|\) and of net field \(|\sum_i h_i|\) at the Corner and Empirical cells, on 400 fresh
instances (seeds 2000--2199 per cell), with Holm correction and the confirmation rule \(\rho\ge0.2\) and \(p_\text{Holm}<0.05\). The \(\rho\ge0.2\) threshold was set, before collecting the held-out
data, as the smallest effect size the exploratory 240-instance pilot could in principle have shown reliably; it is a conventional small-to-moderate effect-size bound, not a power calculation against
a specific alternative. Magnetization \(|M|\) is computed from the certified ground state; net field \(|\sum_i h_i|\) is computed from the programmed fields alone and needs no ground state at all.
Neither uses annealer data.

\textbf{Table II.} Pre-registered confirmatory tests (\(n=200\) per test). A positive \(\rho\) means that a larger predictor value goes with higher annealer occupation, so low \(|M|\) or low
\(|\sum_i h_i|\) marks failure.

\iftwocolumnbuild\onecolumngrid\fi
{\def\LTcaptype{none} 
{\footnotesize\setlength{\tabcolsep}{4pt}\begin{longtable}[]{@{}lllrlrl@{}}
\toprule\noalign{}
Test & Cell & Predictor & \(\rho\) & 95\% CI & Holm \(p\) & Verdict \\
\midrule\noalign{}
\endhead
\bottomrule\noalign{}
\endlastfoot
H1 & Empirical & \(|M|\) & +0.281 & {[}+0.15, +0.40{]} & \(1.1\times10^{-4}\) & confirmed \\
H2 & Corner & \(|M|\) & +0.138 & {[}−0.01, +0.28{]} & \(2.6\times10^{-2}\) & not confirmed \\
H3 & Empirical & \(|\sum h_i|\) & +0.179 & {[}+0.04, +0.31{]} & \(1.1\times10^{-2}\) & not confirmed \\
H4 & Corner & \(|\sum h_i|\) & +0.220 & {[}+0.08, +0.36{]} & \(2.6\times10^{-3}\) & confirmed \\
\end{longtable}}
}
\iftwocolumnbuild\twocolumngrid\fi

Two of four tests confirm, each at a different cell: \(|M|\) predicts failure at the calibrated Empirical cell and net field at the low-field Corner cell (Table II, Fig. 5). At Empirical, the \(|M|\) association
survives conditioning on SA occupation (partial \(\rho=0.318\), \(p=2.4\times10^{-6}\)), so it is not merely a proxy for instances that are hard for every sampler; at Corner it does not (partial
\(\rho=0.075\)). Discrimination is modest: \(|M|\) separates hard instances (occupation \(<5\%\)) with AUC 0.64 at Empirical and 0.59 at Corner (threshold-free); net field discriminates
comparably, AUC 0.642 {[}0.558, 0.719{]} at Corner, where it is pre-registered and confirmed, and 0.577 {[}0.488, 0.665{]} at Empirical, where it is not. To our knowledge, this is the
first held-out, pre-registered test of an \emph{a priori} physical predictor of annealer ground-state failure on embedded instances.

An AUC near 0.64 is a real association, but it does not by itself say what the predictor is worth operationally. We evaluated a concrete triage policy on the same 400 instances: route the bottom
10/20/30\% by the predictor (low \(|M|\) or low net field, the predicted-hardest) to SA instead of the QPU, keeping the rest on the QPU as usual, and compare the realized population-mean occupation
against two baselines --- no triage at all, and triaging a \emph{random} subset of the same size (the fair benchmark a predictor must beat, not just doing nothing). Triage beats doing nothing
substantially (e.g.\ at Empirical, +9.4 percentage points in mean occupation at a 20\% triage fraction, which also removes that fraction of instances from the QPU queue entirely), but the gain over
the random-triage baseline of the same size is small and inconsistent in sign across cells and fractions (\(-0.6\) to \(+0.7\) percentage points). Most of the apparent benefit comes from the fact that
moving \emph{any} instances from QPU to SA tends to help (Sec. III A finds QPU below SA in 392 of 400 instances), not from the predictor's ranking ability specifically. At this AUC, the physical
predictors are statistically real, pre-registered associations, but on their own a weak basis for an operational triage policy.

Two covariates could confound these associations. First, autoscaling: the solver's chain-strength divisor (mean 1.684, SD 0.032 across the 400 instances, essentially constant rather than
instance-dependent) sets the realized hardware scale of the weakest programmed bonds (the tie-strength tier, \(|J|=0.10\)) to about 3\% of the hardware's \(J\) range of 2 (the same convention used in
Sec. III D), so that a 1\% ICE offset is a comparatively large, \(\sim\)34\% relative error on those specific bonds. Neither chain strength nor the realized weak-bond fraction (mean 20.1\%, range 16.3--24.7\%) correlates with occupation on
its own (Spearman \(p=0.56\)--0.95), and the two confirmed predictors survive joint conditioning on both: at Empirical, partial \(\rho=0.303\) (\(p=1.3\times10^{-5}\)), if anything higher than the raw
\(\rho=0.281\); at Corner, partial \(\rho=0.147\) (\(p=0.038\)), weaker than the raw \(\rho=0.220\) but in the same direction. Second, embedding quality: conditioning instead on chain length and
chain-break fraction, by the same method used for the cross-processor regression of Sec. III F, leaves \(|M|\) at Empirical essentially unchanged (partial \(\rho=0.255\), \(p=2.7\times10^{-4}\)) but weakens
net field at Corner below the pre-registered confirmation threshold (partial \(\rho=0.123\), \(p=0.082\)). Chain breaks are rare in this regime (mean fraction 0.0014 at Empirical, 0.00027 at Corner; broken chains are
resolved by majority vote, the Ocean SDK default), so this is a genuine covariate effect rather than an artifact of broken chains: the net-field predictor at Corner, already the weaker of the two
confirmed tests, is the more fragile of the two to embedding covariates.

\textbf{A dedicated replication of the net-field predictor.} To determine whether this fragility reflects a true chain-length confound or merely the power lost to conditioning on three correlated
embedding covariates at \(n=200\), we pre-registered and ran a dedicated follow-up: 300 fresh Corner-cell instances, disjoint from every other sample in this study, each with four embeddings selected
-- from eight minorminer candidates per instance -- for mean chain length closest to a fixed target (4.6, this cell's historical mean on this chip), rather than accepted as found. This decouples
chain length from the instance's own net field by construction rather than by statistical adjustment; the design succeeded (Spearman \(\rho=0.088\), \(p=0.128\), between realized chain length and net
field across the 300 instances, against a pre-registered validity threshold of \(|\rho|<0.15\); chain length landed at \(4.59\pm0.11\)). All 300 instances were submitted to the same processor (224.9~s
total QPU access time). The result is not the ambiguity the design was built to resolve: partial \(\rho(Y,\,\text{net field}\mid\text{SA},\,\text{chain-break fraction},\,\text{chain strength})=-0.006\)
(\(p=0.92\), bootstrap 95\% CI \([-0.12,\,0.11]\)) --- far below the original 200-instance value under the same conditioning set reported above (partial \(\rho=0.123\), \(p=0.082\)) --- and partial
\(\rho\) given SA occupation alone is likewise essentially zero (\(-0.0003\)), not merely reduced. The verdict is simpler than the partial-correlation story above already makes it: H4's own pre-registered
rule required \(\rho\ge0.2\) (SM Table S6), and the raw, unconditioned \(\rho=0.159\) (\(p=0.006\), still significant in direction) already falls short of that threshold on its own, before any conditioning
is applied. That raw association is, in any case, fully accounted for by net field's own correlation with SA occupation in this
sample (\(\rho=0.278\), \(p=1\times10^{-6}\), itself higher than in the original 200-instance set, \(\rho=0.168\)); net field adds nothing to SA's own prediction of occupation once that correlation is
removed. Net field at Corner should therefore be read as an initially promising, pre-registered association that failed a larger, purpose-built, confound-controlled replication -- not as a confirmed
predictor that is merely fragile.

Is this physical predictability a meaningful advance over a simple baseline built from cheap, pre-hardware instance features, in the style of prior machine-learning approaches to this problem
\cite{ref8,ref9}? We fit a logistic-regression baseline on features that need no hardware access and no solved ground state (degree mean and variance, mean \(|J|\), antiferromagnetic-bond fraction,
mean and variance of \(|h|\), edge count), trained only on the same 240-instance exploratory set used to select \(|M|\) and net field, and evaluated, with no further fitting, on the same 400 held-out
instances used for Table II. The answer differs by cell. At Empirical, the cheap-feature baseline, driven almost entirely by antiferromagnetic fraction, reaches AUC 0.661 {[}0.575, 0.744{]}, statistically
indistinguishable from \(|M|\)'s 0.64: a simple frustration proxy matches the pre-registered physical predictor here. At Corner, the same baseline reaches only AUC 0.522 {[}0.436, 0.611{]}, well below net
field's 0.642 with little CI overlap: net field showed an advantage over cheap features in this held-out AUC test --- but net field is also the predictor that failed the larger, confound-controlled
replication above, so this advantage does not survive either. A combined model (cheap features plus \(|M|\) and net field together) does
not outperform the best single predictor at either cell, consistent with overfitting on 120 training instances per cell once feature count grows. Taken together, no pre-registered physical predictor
demonstrates a replicated advantage over a cheap, hardware-agnostic frustration proxy at either cell: at Empirical the frustration proxy already matches \(|M|\), and at Corner the one predictor that
beat it, net field, is the one that failed to replicate.

This section's results should be kept at three distinct levels, since they do not all hold to the same standard: \emph{statistical significance} (a nonzero, held-out correlation on the original
400-instance sample, met by both \(|M|\) at Empirical and net field at Corner); \emph{independent replication} (met only by \(|M|\); net field's Corner association did not survive a dedicated,
confound-controlled, 300-instance follow-up); and \emph{practical prediction} (met by neither, since both lose their apparent edge once compared against a random-triage or cheap-feature baseline of
the same effect size). A predictor can satisfy the first without the second or third, and conflating the three overstates what Table II alone supports.

\begin{figure}[tbp]
\centering
\includegraphics[max width=\linewidth]{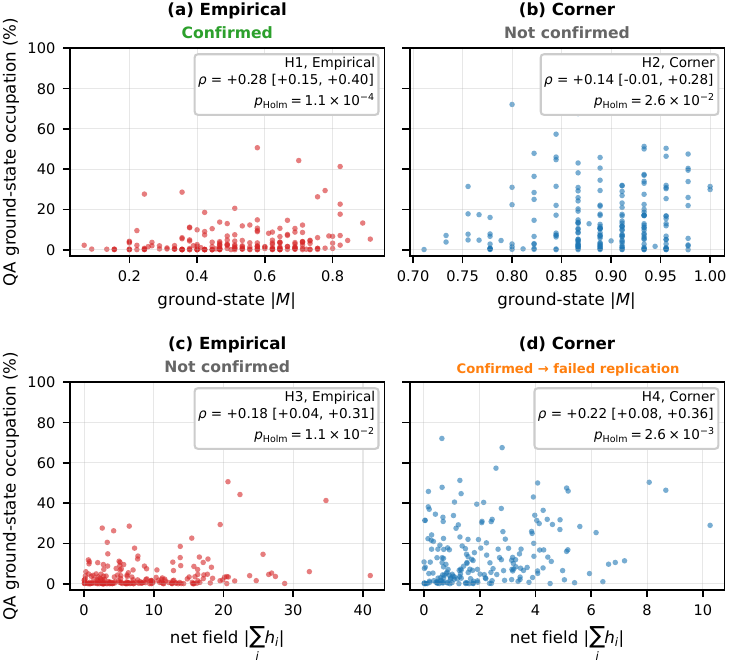}
\caption{Pre-registered confirmatory tests on the 400 held-out instances: QA occupation versus (a, b) ground-state magnetization \(|M|\) and (c, d) net field \(|\sum_i h_i|\), at the Empirical cell (a, c) and the Corner cell (b,
d). Insets give the one-sided Spearman \(\rho\) with its bootstrap 95\% confidence interval, the Holm-adjusted \(p\) value, and the verdict under the pre-registered rule (\(\rho\ge0.2\) and
\(p_\text{Holm}<0.05\)), evaluated on this 400-instance sample. Panel (d)'s net-field test at Corner (H4) met that rule here but failed a larger, dedicated, confound-controlled replication reported
in Sec. III C; its label reflects both outcomes rather than only the status at the time of this figure's own sample.}
\end{figure}

\subsection{Static control-error perturbations reproduce part of the instance ranking}\label{static-control-error-perturbations-reproduce-part-of-the-instance-ranking}

To test whether static miscalibration can account for the instance ranking, we modeled ICE as a \emph{fixed} offset per submission rather than noise redrawn every read, the better analogue of how ICE
arises. Flux offsets and quantization errors are properties of particular qubits under a given calibration, so they are approximately constant across the reads of one submission; such persistent, qubit-specific biases have been experimentally characterized and, once corrected, shown to improve annealer performance \cite{ref39}, though that does not by itself establish that the biases measured here cause the instance-dependent failure under study. For each of the 400
held-out instances and each noise level (1\%, 2\% and 4\% of the programmable \(h\) and \(J\) ranges, applied in problem units), we drew five fixed Gaussian offsets. We solved each perturbed problem
with 200 SA reads and recorded the fraction of reads at the \emph{unperturbed} certified ground state. This synthetic occupation was then correlated with the annealer's measured occupation on the same
instance. The solver auto-scales each embedded problem by its chain strength (mean 1.7 here), so these offsets correspond to about 0.6\%, 1.2\% and 2.4\% of the hardware's ranges. Concretely, at the 1\% level
\(\sigma_h=0.12\) and \(\sigma_J=0.02\) in problem units, equivalent to \(\approx0.071\) and \(\approx0.012\) hardware units after the chain-strength divisor, against hardware ranges of 12 (\(h\))
and 2 (\(J\)); both land at \(\approx0.6\%\) of their respective range because the same divisor is applied to both, not because \(h\) and \(J\) carry the same absolute offset. The 2\% and 4\% levels
scale these numbers linearly.

The model perturbs the instance and then runs SA, so its output inherits SA's own instance ranking. Unperturbed SA occupation alone predicts annealer occupation with \(\rho=0.54\) {[}0.43, 0.63{]} at
Corner and \(0.50\) {[}0.38, 0.61{]} at Empirical. Table III therefore also reports the model's partial correlation with the annealer given SA occupation, computed on ranks.

\textbf{Table III.} Spearman correlation between synthetic static-error occupation and annealer occupation (\(n=200\) per row), and the partial correlation given SA occupation (bootstrap 95\% CI).
Offsets are in problem units, with the corresponding share of the hardware range (hw.) in parentheses.

{\def\LTcaptype{none} 
{\footnotesize\setlength{\tabcolsep}{4pt}\begin{longtable}[]{@{}lrrrlr@{}}
\toprule\noalign{}
Cell & Offset (hw.) & \multicolumn{1}{c}{\(\rho\)(model, QA)} & \multicolumn{1}{c}{\(\rho\)(model, SA)} & \multicolumn{1}{c}{Partial \(\rho\)} & \multicolumn{1}{c}{\(p\)} \\
\midrule\noalign{}
\endhead
\bottomrule\noalign{}
\endlastfoot
Corner & 1\% (0.6\%) & +0.557 & +0.761 & +0.27 {[}+0.14, +0.39{]} & \(1.0\times10^{-4}\) \\
Corner & 2\% (1.2\%) & +0.413 & +0.444 & +0.23 {[}+0.09, +0.37{]} & \(1.1\times10^{-3}\) \\
Corner & 4\% (2.4\%) & +0.321 & +0.231 & +0.24 {[}+0.10, +0.37{]} & \(6.4\times10^{-4}\) \\
Empirical & 1\% (0.6\%) & +0.384 & +0.728 & +0.03 {[}−0.11, +0.17{]} & 0.69 \\
Empirical & 2\% (1.2\%) & +0.406 & +0.546 & +0.18 {[}+0.04, +0.31{]} & 0.011 \\
Empirical & 4\% (2.4\%) & +0.094 & +0.270 & −0.05 {[}−0.20, +0.10{]} & 0.47 \\
\end{longtable}}
}

The three noise levels of Table III are a sensitivity sweep over one classical model, run on the same 400 held-out instances as Table II, not three independent confirmatory tests; they were not
pre-registered and carry no multiplicity correction beyond what Table II's own Holm correction already covers for the held-out sample. At the low-field Corner cell, the static-error model carries
information about the annealer's instance ranking beyond classical hardness at every noise level (partial \(\rho=0.23\)--\(0.27\)). At the
calibrated Empirical cell, most of its correlation is shared with SA occupation. A residual association remains only at the 2\% level, and none at 1\% or 4\%. The relevant offsets (0.6--1.2\% of the hardware
range) lie at or below the documented ICE scale of about 1\% \cite{ref1}. At 2.4\% the Empirical correlation vanishes, consistent with perturbations large enough to reorder low-lying levels for most
instances. A per-read (resampled) version of the same model, tested earlier on six instances, gave only directional support (SM S6).

\begin{figure}[tbp]
\centering
\includegraphics[max width=\linewidth]{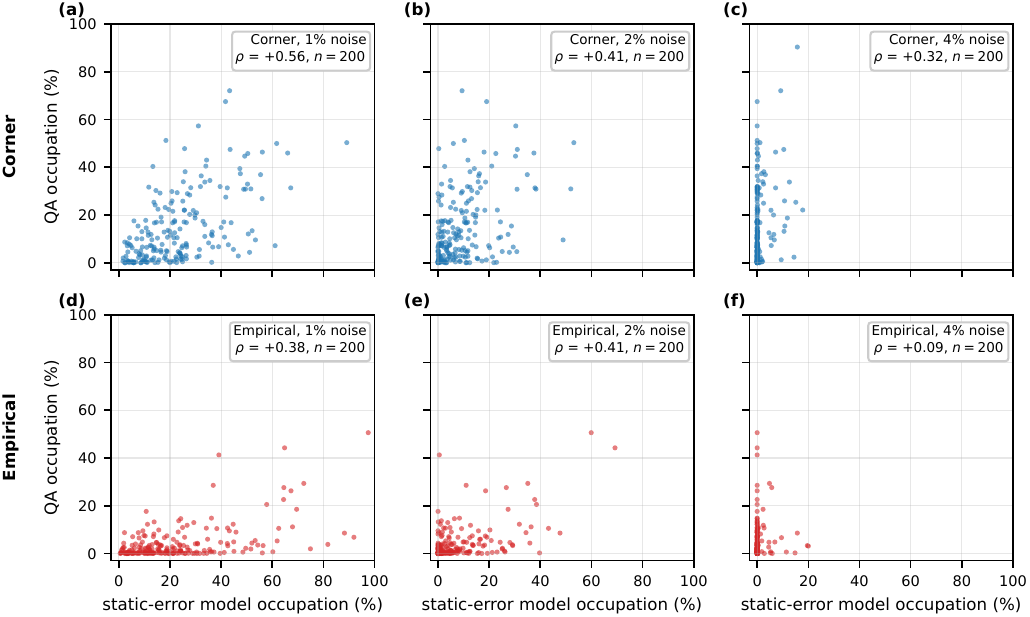}
\caption{QA occupation versus ground-state occupation under the static control-error model, for the Corner cell (a--c) and the Empirical cell (d--f), at offset standard deviations of 1\%, 2\% and 4\% of the programmable range; 200
instances per panel. Spearman coefficients as in Table III.}
\end{figure}

\subsection{Hardware diagnostics of the competing states}\label{hardware-diagnostics-of-the-competing-states}

Several diagnostics, first run on a single realization \((10,0.25,0.20)\), seed 42 (occupation 0.6\%), and extended to scale below, constrain alternatives to a static-error explanation; full discussion
is in SM Sec. S14. Neither a 25\(\times\) or 100\(\times\) slower anneal, a 100 µs mid-anneal pause, nor halving or doubling the chain strength recovered the ground state from this instance's competing
state (Fig. 7(a); consistent with the small pooled anneal-time effect of Sec. III B and the sweep result that weaker chains help). Reverse annealing from the competing state (Table IV, Fig. 7(b)) --
following the use of reverse annealing to map transitions between Ising configurations \cite{ref34}, and its use to search cyclically for deep low-energy states in large spin glasses \cite{ref48} -- showed the same one-way asymmetry later confirmed at scale (Sec. III F): full recovery from a
near-degenerate competing state (Hamming distance 1) but none at all from the other kind (raw Hamming distance 76--83 of 90) -- which, despite its large raw distance, sits only 7--14 spins from
the \emph{flipped} ground state, so we call it a ``near-flip'' rather than ``distant'' trap throughout (the \(Z_2\)-symmetry/folding argument behind this is given in full in Sec. III F) -- across five realizations at this cell point. This is not proof of a particular mechanism \cite{ref23}.

\textbf{Table IV.} Reverse annealing from the certified ground state (GS) and from the competing state (CS), averaged over realizations at each competing-state regime (near-flip trap, \(n=2\); near-degenerate, \(n=3\); see text).

{\def\LTcaptype{none} 
{\footnotesize\setlength{\tabcolsep}{4pt}\begin{longtable}[]{@{}llrrrr@{}}
\toprule\noalign{}
Competing state & Initial state & \(s_\text{target}\) & GS & CS & Other \\
\midrule\noalign{}
\endhead
\bottomrule\noalign{}
\endlastfoot
Near-flip trap & GS & 0.75 & 81.1\% & 0.0\% & 18.9\% \\
Near-flip trap & GS & 0.55 & 71.5\% & 0.0\% & 28.5\% \\
Near-flip trap & GS & 0.35 & 38.5\% & 0.0\% & 61.5\% \\
Near-flip trap & CS & 0.75 & 0.0\% & 69.3\% & 30.7\% \\
Near-flip trap & CS & 0.55 & 0.0\% & 59.2\% & 40.8\% \\
Near-flip trap & CS & 0.35 & 0.0\% & 10.5\% & 89.5\% \\
Near-degenerate & GS & 0.75 & 91.4\% & 8.6\% & 0.1\% \\
Near-degenerate & GS & 0.55 & 87.5\% & 11.6\% & 0.8\% \\
Near-degenerate & GS & 0.35 & 49.3\% & 18.5\% & 32.2\% \\
Near-degenerate & CS & 0.75 & 12.5\% & 87.4\% & 0.0\% \\
Near-degenerate & CS & 0.55 & 17.5\% & 81.7\% & 0.8\% \\
Near-degenerate & CS & 0.35 & 25.2\% & 44.3\% & 30.5\% \\
\end{longtable}}
}

\emph{Embedding and gauge dependence.} Six of eight fresh embeddings on disjoint physical qubits converged on the same near-flip competing-state region (Fig. 7(c); full detail in SM Sec. S14), so the
competing state is tied to the encoded problem, not to one chain layout, although spatially local crosstalk among co-located embeddings is not excluded (Sec. IV). Gauge averaging raised this single
instance's occupation about tenfold but did not remove the competing state (SM S7); at scale it does not help at all (SM Fig. S13), null at Corner (paired Wilcoxon \(p=0.14\)) and slightly harmful at
Empirical (\(p=0.038\), mean \(-0.6\) pp) over 120 instances per cell.

\begin{figure}[tbp]
\centering
\includegraphics[max width=\linewidth]{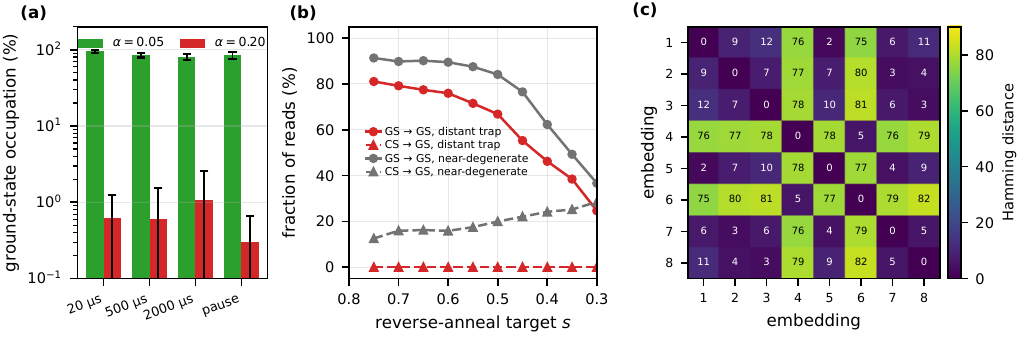}
\caption{Illustrative, not confirmatory: diagnostics at the low-field corner (\(N=90\), \(m=10\), \(r_h=0.25\)), motivating the scaled 12-instance test of Table V. (a) and (c) use a single instance
(seed 42), combining the \(\alpha=0.05\) and \(\alpha=0.20\) conditions in (a). (a) Ground-state occupation at \(\alpha=0.05\) and \(\alpha=0.20\) under the default 20 µs anneal,
500 µs and 2,000 µs anneals, and a 100 µs mid-anneal pause; error bars, standard deviation across embeddings. (b) Reverse annealing at \(\alpha=0.20\), averaged over five realizations at this cell point
(two landing on a near-flip competing state, three on a near-degenerate one; Table IV): fraction of reads that remain in, or reach, the ground state (GS) versus the target \(s\), starting from the
ground state (GS \(\to\) GS) or from the competing state (CS \(\to\) GS). (c) Pairwise Hamming distance between the modal low-energy states returned by eight independent embeddings, seed 42.}
\end{figure}

\emph{Generalization to held-out hard and easy instances.} To test whether the seed-42 picture holds beyond one instance, we repeated the diagnostics on 12 instances from the held-out sample at the
calibrated Empirical cell (\(N=90\), \(m=10\), \(\alpha=0.20\), \(r_h=1\)). The selection rule was fixed before any further hardware contact: the six ``hard'' instances are those with zero annealer occupation
in all four held-out embeddings and the highest SA occupation (37--79\%, so the failure is specific to the annealer), and the six ``easy'' instances are those with the highest annealer occupation
(28--51\%). Each was run on two of its held-out embeddings under 13 conditions (1,000 reads each; 400 at 2,000 µs), 312 submissions in total (Table V).

\textbf{Table V.} Mean ground-state fraction for six hard and six easy instances at the Empirical cell (two embeddings each). CS, the dominant non-ground state of the 20 µs forward anneal on the same embedding.
Mann--Whitney \(p\) compares per-instance means (6 vs 6; the smallest attainable two-sided \(p\) is 0.0022).

{\def\LTcaptype{none} 
{\footnotesize\setlength{\tabcolsep}{4pt}\begin{tabular}{@{}lrrr@{}}
\toprule
Condition & Hard & Easy & \(p\) \\
\midrule
Forward, 20 µs & 0.000 & 0.308 & 0.004 \\
Forward, 200 µs & 0.000 & 0.386 & 0.004 \\
Forward, 2,000 µs & 0.000 & 0.308 & 0.004 \\
Forward, chain strength ×0.5 & 0.000 & 0.372 & 0.017 \\
Forward, chain strength ×2 & 0.000 & 0.036 & 0.004 \\
Reverse from GS, \(s=0.45\) & 0.813 & 0.739 & 0.39 \\
Reverse from GS, \(s=0.35\) & 0.480 & 0.512 & 0.82 \\
Reverse from GS, \(s=0.30\) & 0.157 & 0.406 & 0.015 \\
Reverse from GS, \(s=0.26\) & 0.009 & 0.332 & 0.002 \\
Reverse from CS, \(s=0.26\)--\(0.45\) & 0.000 & 0.23--0.29 & \(\le0.01\) \\
\bottomrule
\end{tabular}}
}

The hard instances never reached the ground state in any forward condition: neither a 100\(\times\) longer anneal nor halving or doubling the chain strength produced a single ground-state read. Their
dominant competing state lies a median 10 spins (range 4--46) and 1.53 energy units above the ground state, against 1 spin and 0.31 for the easy instances. Reverse anneals reproduce the seed-42
barrier: started from the competing state, the hard instances returned to the ground state in none of 48 runs at any \(s_\text{target}\) from 0.26 to 0.45, whereas easy instances did so in 23--29\% of
reads. Started from the ground state, hard and easy instances are equally stable at \(s\ge0.35\); at \(s\le0.30\) the hard instances lose it almost completely (1\% retained at \(s=0.26\)), while the
easy instances retain a third. The hard instances' competing states are closer to the ground state than seed 42's (median 10 versus 74 spins), so seed 42's own raw distance does not generalize, but the asymmetric, one-way recovery pattern does --
in the schedule-dependent dynamical-trapping sense established in Sec. III F, not as a demonstrated, unconditional barrier.

\emph{Chip location.} To test the crosstalk alternative directly, we partitioned the chip into four disjoint quadrants and embedded each of the 12 instances once in each quadrant (48 submissions).
Instance identity explains 94\% of the variance in ground-state fraction and quadrant 0.9\% (Friedman test across quadrants, \(p=0.39\)); this split is inflated by the extreme, bimodal selection
of the 12 instances (Y=0 throughout for hard, Y=0.28--0.51 for easy; see the selection rule above) and should not be read as the variance split on a representative sample. Hard instances failed in every quadrant (maximum 1.0\%
ground-state reads; 16 of 24 instance--quadrant pairs exactly zero), and easy instances succeeded in every quadrant (4.7--92\%). Failure is a property of the instance, not of the chip region on which
it is placed.

\emph{Static-error test on the same instances (full detail in SM Sec. S14).} If the hard instances fail because the realized Hamiltonian differs from the programmed one, their certified ground state
should be more fragile to fixed offsets of ICE size than that of the easy instances. Re-solving each of the 12 instances exactly under 40 fixed Gaussian offsets at 1\% and 2\% of the hardware ranges
(960 Gurobi solves) confirms this: at 1\%, the certified ground state was no longer optimal in 92\% of draws for hard instances against 63\% for easy ones (\(p=0.005\)), pushing the optimum a median
5 spins away for hard instances against 1 for easy ones. The simplest version of the account fails, however: the probability that the competing state specifically falls below the ground state under
the same offsets is identical for both groups (0.25, \(p=0.41\)), and the new optimum was rarely the specific competing state the hardware returned (3\% of draws for hard instances). Static errors
thus make the hard instances' ground states fragile and push the optimum several spins away, but the trapping asymmetry of Table V points to a collective competing state the dynamics rarely leaves at
this schedule, which is not a property of the static error alone.

\emph{Sparse-graph check (exploratory).} Neither predictor was pre-registered at \(m=2\). We therefore ran 50 fresh instances at each of two sparse cells, \(N=90\), \(\alpha=0.20\), \(m=2\), with
\(r_h=1\) (E2) and \(r_h=0.25\) (C2). Each instance used four embeddings × 1,000 reads, and all 100 reference ground states were certified. The annealer-to-SA occupation ratio matches the size and
field sweeps (Figs. 2--3; Table VI).

\textbf{Table VI.} Sparse-graph check (\(N=90\), \(m=2\), \(\alpha=0.20\); \(n=50\) instances per cell).

{\def\LTcaptype{none} 
{\footnotesize\setlength{\tabcolsep}{4pt}\begin{longtable}[]{@{}lrrrr@{}}
\toprule\noalign{}
Cell & \(r_h\) & QA mean & SA mean & Ratio \\
\midrule\noalign{}
\endhead
\bottomrule\noalign{}
\endlastfoot
E2 & 1 & 32\% & 44\% & 0.73 \\
C2 & 0.25 & 12\% & 10\% & 1.14 \\
\end{longtable}}
}

C2 is at parity: QA occupation exceeds SA occupation in 23 of 50 instances. SA occupation is the strongest correlate of QA occupation at both cells (Spearman \(\rho=0.70\) and \(0.57\),
\(p<10^{-4}\)).

The magnetization predictor that holds at the dense calibrated Empirical cell does not carry over to the sparse E2 (\(\rho=0.00\)). At E2 the energy gap to the first excited state is the better correlate
instead (\(\rho=0.61\)). At C2, \(|M|\) and \(|\sum_i h_i|\) give weak positive correlations: \(\rho=0.30\) (\(p=0.037\)) and \(\rho=0.29\) (\(p=0.042\)), uncorrected. These are exploratory and
consistent with the regime dependence of Sec. III C.

\subsection{Robustness across days, processors, gauges and system size}\label{robustness}

The results above come mostly from one processor at one calibration. To test whether they are properties of the instances rather than of one chip at one moment, we ran a further pre-registered
campaign (Q1--Q7 below) spanning four days, three processors, a new graph family and a new density, spin-reversal gauges and a second size-scaling run. A later, separately pre-registered addendum
(Q8) adds a single further prediction: a dedicated, chain-length-matched replication of the weaker of the two preregistered predictors (net field at Corner; Sec. III C), run once that predictor's
fragility to embedding covariates was noticed in the Q1--Q7 campaign. Together with the predictor test of Sec. III C, this brings
the study's held-out, pre-registered predictions to 87, of which 61 held (45 of 71 once the near-certain-by-design stability checks below are excluded; Table VII); the full prediction-by-prediction
ledger, with each prediction's registration date and confirmation rule, is given in the Supplemental Material (Table S6).

\textbf{Day-to-day stability.} The same 400 held-out instances and embeddings were resubmitted on four further days (14, 15, 15 and 15 days after the original run, the last three on one calibration
day by the author's decision, logged before any data were seen). The instance ranking is stable throughout: Spearman \(\rho\) with the original run is 0.989--0.990 at the Empirical cell and 0.994--0.995 at Corner on
every day, hard/easy agreement is 95--98\%, and the median change in occupation stays within \(\pm0.6\) percentage points with no trend across days. All 16 pre-registered predictions (two per day,
two cells) hold.

\textbf{Cross-processor transfer.} The 400 instances of the Corner and Empirical cells were resubmitted on two Pegasus-topology processors (\texttt{Advantage\_system4}, \texttt{Advantage\_system6}; 400 instances
\(\times\) 4 fresh embeddings \(\times\) 3,000 reads each). The annealer falls below SA on every one of 1,600 instance--processor pairs. The per-instance ranking transfers strongly to
\texttt{Advantage\_system4} (Spearman \(\rho=0.60\) at Empirical, \(0.88\) at Corner, both \(p<10^{-9}\)) and more weakly to \texttt{Advantage\_system6} (\(\rho=0.23\), \(0.18\)), which runs closer to the
occupation floor (Fig. 8a--b). Classical hardness (SA occupation) predicts the QPU outcome on every chip (\(\rho=0.44\)--\(0.54\)) and is therefore the processor-independent component of the ranking.
Rank regression that removes SA occupation and each processor's chain-break rate and length leaves a residual agreement between Zephyr and \texttt{Advantage\_system4} (\(\rho=0.41\) at Empirical, \(0.82\) at
Corner), evidence for an instance-specific effect beyond classical hardness and chain quality that both chips share; \texttt{Advantage\_system6} shares none of this residual, and even reverses sign on
\(|M|\) at Empirical, consistent with its own, different static-error pattern (see the measured-offset analysis below). On a separate run (Q7), the size-scaling collapse of Fig. 2c and its per-instance
ranking transfer reproduce decisively on \texttt{Advantage\_system4} (Fig. 8c), as already reported in Sec. III B. The QA-below-SA finding itself
holds cleanly at the two middle sizes (\(N=90\), 98\% of instances; \(N=180\), 98\%) and is not testable at the extremes for two understood, opposite reasons: at \(N=30\) the annealer genuinely
outperforms SA on 14\% of instances (a ceiling effect, not a ranking failure), and at \(N=360\) both solvers return exactly zero occupation on 19 of 60 instances (a floor effect that the strict
below-SA test cannot score).

\begin{figure}[tbp]
\centering
\includegraphics[max width=\linewidth]{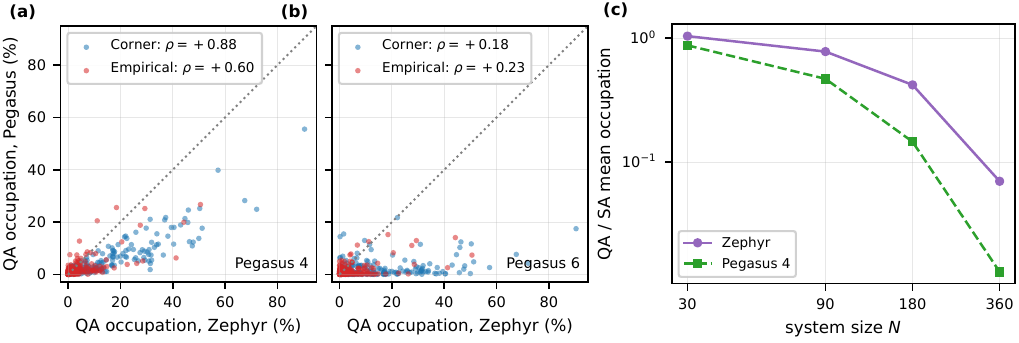}
\caption{Cross-processor robustness (Q2, Q7). (a)--(b) Per-instance QA occupation on Zephyr versus the same instance on two Pegasus processors, the Corner and Empirical cells (200 instances per cell; Spearman
\(\rho\) in the legend). Dotted line, equal occupation. (c) QA/SA mean occupation ratio versus system size \(N\) at \(m=2\) (pooled over \(r_h\)), Zephyr (Fig. 2c) versus
\texttt{Advantage\_system4} (Q7); 30 instances per point.}
\end{figure}

\textbf{Transfer to new cells, a new graph family and gauges.} In three further cells, including a new graph family (an Erd\H{o}s--R\'enyi \(G(n,m)\) random graph with edges placed uniformly at random, matched to the baseline graph's edge count rather than its degree sequence, cell ``ER'') and a sparser Barab\'asi--Albert
density (cell ``E5'', \(m=5\)), the annealer again falls below SA in \(\ge99\%\) of 400 fresh instances and classical hardness again predicts the outcome (\(\rho=0.44\)--\(0.61\)). The magnetization
predictor transfers, with its pre-registered sign, to the new graph family ER (\(\rho=0.20\), \(p<10^{-3}\)) but not to the sparser E5 (\(\rho=0.04\), \(p=0.20\)): field share alone does not
determine where it works, and density appears to matter as well. Spin-reversal gauges (eight per instance, both cells, both Zephyr and \texttt{Advantage\_system4}) show that a static, instance-specific
bias is widespread: occupation depends significantly on the gauge in 44--97\% of instances on both chips, and the instances the static-error model of Sec. III D calls most fragile are the ones whose
occupation swings most with the gauge (\(\rho=0.26\)--\(0.43\), all \(p<10^{-3}\)), on both chips. This is a hardware measurement, not a classical model, of the static-error account. It also sharpens
the regime split: at the calibrated Empirical cell, 89--94\% of the hard instances on Zephyr and \texttt{Advantage\_system4} stay hard under every gauge, so their failure is intrinsic to the instance; at the
low-field Corner cell, 45\% of Zephyr's hard instances escape under some gauge (a largely removable static bias), while 84\% stay hard on \texttt{Advantage\_system4} (largely intrinsic there). The same
split therefore runs in different directions on the two processors.

\emph{Landscape geometry predicts an instance's own occupation.} The remainder of this section turns from cross-day/processor/cell/gauge robustness to a different question: whether the hardware's own
observed competing state says anything about an instance's occupation, beyond what is already known before running it.

\textbf{Trapping generalizes.} A fresh sample of 24 hard and 24 easy instances (disjoint from the 6+6 of Sec. III E, selected by the same pre-registered rule) was tested for the same reverse-annealing
asymmetry, 8 reverse-anneal runs per instance. On Zephyr, the hard instances return to the ground state from their competing state in 6 of 192 runs (3\%) against 120 of 192 (63\%) for easy instances;
on \texttt{Advantage\_system4} the same asymmetry holds in direction but is less extreme (44\% vs.\ 86\% of runs), consistent with noisier reverse anneals and longer chains on that chip. Because the 8
runs per instance share one landscape, embedding and competing state, we treat the instance, not the run, as the unit of inference: at the instance level (per-instance recovery fraction over its 8
runs, \(n=24\) vs.\ 24), easy instances recover more often than hard ones on both chips (Mann--Whitney, Zephyr \(p=9\times10^{-10}\); \texttt{Advantage\_system4} \(p=8\times10^{-6}\)), and an
instance-clustered bootstrap on the pooled hard/easy recovery-rate gap (resampling instances, not runs, 10,000 draws) excludes zero on both chips (Zephyr: gap 0.59, 95\% CI [0.48, 0.70];
\texttt{Advantage\_system4}: gap 0.42, 95\% CI [0.28, 0.57]) --- far less extreme than a run-pooled Fisher test on the same counts (\(p=4\times10^{-40}\) and \(10^{-18}\), respectively, which overstate
precision by treating correlated runs as independent) but still decisive. Figure 9 shows this instance-level recovery-fraction distribution directly, hard versus easy, on both chips. In both cases the competing state lies substantially farther from the ground state for hard than for easy instances (Zephyr:
median 12.5 vs.\ 2 spins; Pegasus: 6.5 vs.\ 2). This is now a 48-instance, two-processor measurement rather than the single-instance and 6+6-instance results of Sec. III E, and it supports calling the
asymmetry a \emph{schedule-dependent dynamical trapping signature}, strongly direction-consistent across instances and chips, rather than a proven, unconditional one-way barrier.

\begin{figure}[tbp]
\centering
\includegraphics[max width=\linewidth]{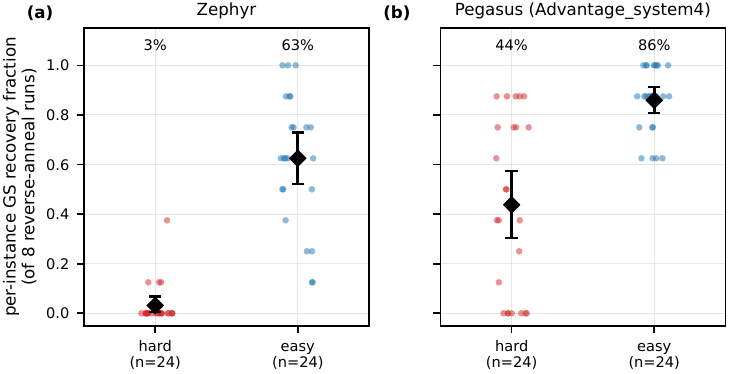}
\caption{Per-instance ground-state recovery fraction under reverse annealing from the competing state (8 runs per instance), hard versus easy, on (a) Zephyr and (b) \texttt{Advantage\_system4}
(Sec. III F, ``Trapping generalizes''). Each point is one instance's own recovery fraction; the black diamond and error bars are the group mean and its instance-clustered bootstrap 95\% CI (10,000
resamples). Percentages above each group are the same run-pooled figures reported in the text.}
\end{figure}

Because the competing state lies farther from the ground state for hard than for easy instances, the recovery asymmetry could in principle be a generic distance effect rather than anything specific
to hardness: that reverse-anneal success depends strongly on the Hamming distance between the initial state and the ground state has been established since early use of the technique \cite{ref44},
so without controlling for it the asymmetry we report would be the expected outcome of the distance gap alone, not evidence of hardness-specific trapping. It is not reducible to distance alone, though distance also matters independently. Here ``the competing state'' takes two distinct senses that we distinguish explicitly: the \emph{fixed} state used to \emph{initialize} each reverse-anneal run, recorded once per embedded realization from the Sec. III E diagnostic before any reverse-anneal run and shared across that embedding's four runs (two embeddings per instance, eight runs total), and the \emph{per-run outcome} --- that run's own modal non-ground-state read, which in principle could itself be shaped by whether the run recovered. The primary regression therefore conditions on the former, not the latter: a logistic regression of recovery on hard/easy class and each run's fixed initial-state Hamming distance to the ground state, with standard errors clustered by instance (384 runs, 48 instance clusters per chip), finds both terms significant
after conditioning on the other (Zephyr: class coefficient \(-1.75\), cluster-robust \(p=4.1\times10^{-5}\); Hamming coefficient \(-0.61\), cluster-robust \(p=1.3\times10^{-6}\); \texttt{Advantage\_system4}: class
\(-1.03\), \(p=2.4\times10^{-4}\); Hamming \(-0.20\), \(p=1.0\times10^{-7}\)). A secondary, descriptive regression using each run's own post-hoc outcome distance instead gives the same conclusion (Zephyr: class \(-1.67\), \(p=4\times10^{-4}\); Hamming \(-0.58\), \(p=2\times10^{-6}\); \texttt{Advantage\_system4}: class \(-0.84\), \(p=0.0023\); Hamming \(-0.24\), \(p=1\times10^{-7}\)) -- unsurprising given how closely the two distances track each other in practice (Spearman \(\rho=0.99\) on Zephyr, \(0.86\) on \texttt{Advantage\_system4}), but reported only as a robustness check, since the outcome measure is not a clean pre-treatment control. As a descriptive check, matching directly on the fixed initial-state distance shows the same pattern: at Hamming distance 4--10 on Zephyr, hard instances recover in 1.3\% of runs
(\(n=80\) runs, drawn from 11 instances) against 30.6\% for easy instances at the same distance (\(n=36\) runs, 9 instances); on \texttt{Advantage\_system4} the same distance band gives 50.0\% (48
runs, 11 instances) versus 70.0\% (40 runs, 8 instances). This matched comparison is a descriptive robustness check on a coarser distance bin, not an independent confirmation; the inferential
support for the hard/easy gap surviving distance control comes from the clustered regression above, on both chips.

Hamming distance is a configuration-space distance, not a barrier height: a competing state 12 flips away could in principle sit behind a low-energy pathway, while one 2 flips away could be separated
by a high barrier. We therefore estimated an actual energy barrier for all 48 Zephyr Q5 instances: a zero-temperature, least-bottleneck flip path from the ground state to the competing state (at each
step, flip whichever still-differing spin gives the lowest resulting energy, the standard tractable proxy for a minimal-energy path on a landscape too large to search exactly; best of 40 restarts,
half greedy and half random-order as an upper-bound control). Hard instances sit behind a significantly higher barrier than easy ones (median 2.8 vs.\ 0.8 energy units, Mann--Whitney \(p=2\times10^{-4}\)).
This barrier estimate is itself very strongly correlated with Hamming distance in this ensemble (Spearman \(\rho=0.90\)): the sample does not contain the geometrically-close-but-energetically-separated
configurations that would be needed to decouple distance from barrier height, so the two measures corroborate rather than compete here. Given that collinearity, we do not
further decompose their separate contributions (a residual-based partial correlation is numerically unstable at \(\rho=0.90\) and, when attempted, changes sign between the raw and residualized versions
--- itself evidence that it should not be trusted); the barrier estimate supports, rather than supersedes, the distance-based account above, and together they make the dynamical-trapping reading of
Sec. III E--F an estimated energetic property of these instances, not merely a configuration-space one, while still an upper-bound estimate from a heuristic search rather than an exact barrier, and a
purely energetic one: it cannot distinguish an energy barrier from an entropic one, which on other hard optimization problems has been shown to affect classical and quantum algorithms alike \cite{ref46}.
Figure 10 shows this flip path directly, not as a schematic, for one representative hard and one representative easy instance.

\begin{figure}[tbp]
\centering
\includegraphics[max width=\linewidth]{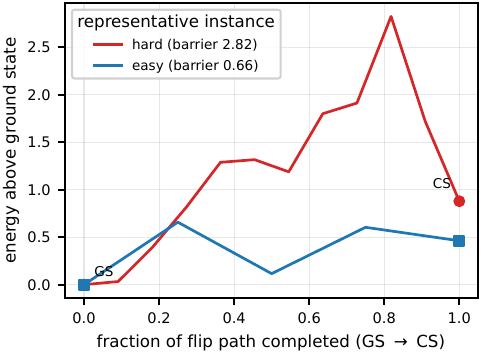}
\caption{One estimated energy path from the ground state (GS) to the competing state (CS), not the full landscape, for one representative hard and one representative easy instance (cell E, each the instance whose barrier is
closest to its class median; Sec. III F), along the same greedy least-bottleneck flip path used for the barrier estimate there. The horizontal axis is the fraction of that configuration-space path
completed, not a physical or continuous coordinate; the path itself is real, reconstructed from each instance's certified ground state and QPU-observed competing state, not schematic. Only the two
endpoints are verified landscape features (GS, the exactly certified global minimum; CS, the hardware's observed read); intermediate points are waypoints of this one heuristic path, not independently
confirmed local minima of the full landscape -- a dip here is a low point relative to its neighbors \emph{on this path only}, since the search minimizes the path's peak energy, not its roughness.}
\end{figure}

Everything above tests whether geometry explains the \emph{recovery asymmetry}; it leaves open whether geometry says anything about the instance's own hardware failure, as opposed to merely
co-occurring with it. We flag explicitly that this predictor sits in a different category from \(|M|\) and net field above: both of those are available \emph{before any hardware run} (one needs the
certified ground state, the other needs only the programmed instance), whereas the competing state (CS) used here is itself observed during QPU sampling, so what follows is a hardware-derived
diagnostic association, not a prospective predictor an experimenter could compute in advance. A classically-estimated proxy for landscape geometry, for instance a search for low-lying local minima
on the programmed instance alone, would be needed to turn this into a prospective test; we did not attempt one and flag it as a direction for future work rather than claim it here. On the same 48 instances, CS Hamming distance predicts the instance's measured QA ground-state occupation \(Y\) beyond both classical hardness (SA occupation) and the confirmed
physical predictor at this cell (\(|M|\)): partial \(\rho=-0.59\) (\(p=9\times10^{-6}\)), a bootstrap-stable estimate (10,000 resamples, 95\% CI \([-0.77,-0.32]\), entirely one-signed). The barrier
estimate shows the same pattern on its own, more weakly (partial \(\rho=-0.41\) given SA and \(|M|\), \(p=0.004\); bootstrap CI \([-0.65,-0.08]\)); because barrier and Hamming distance are themselves
collinear (\(\rho=0.90\)), conditioning on both together is unstable, as above, so this is reported as two separate single-control results, not a joint decomposition. Landscape geometry therefore
carries predictive information about this cell's hardware failure beyond what is already captured by classical hardness and the pre-registered predictor --- a stronger evidentiary status than
``co-occurs with hardness.'' This was first shown on a small, extreme-selected diagnostic subsample (the top and bottom 24 of a larger pool by occupation, not a random held-out sample), so we
repeated it on samples never selected on \(Y\), in two stages that we label explicitly by their evidentiary status: a \emph{retrospective} reanalysis of data collected for another purpose, and a
\emph{prospective} test, its analysis plan fixed before collection, on newly collected data -- not part of the formal, time-stamped pre-registration ledger of Table VII, like T1--T4c above. The retrospective stage uses the full 400-instance Table II held-out set --- the
same 400 instances behind Table II above, collected before this question was asked, reanalyzed rather than newly sampled (200 per cell). Because the competing state and \(Y\) are computed from the same embeddings' reads (justified below), we report the more conservative,
independently-sampled split-embedding estimate as the headline number throughout: partial \(\rho=-0.26\) to \(-0.35\) at Empirical given SA occupation and \(|M|\) (same-embedding estimate \(-0.49\), \(p=2\times10^{-13}\));
\(-0.61\) at Corner given SA occupation and net field --- this cell's own preregistered predictor, not \(|M|\) --- (same-embedding estimate \(-0.72\), \(p=9\times10^{-34}\); conditioning on \(|M|\) at Corner instead gives the same pattern, and conditioning on SA occupation alone, with no physical predictor at all, gives \(-0.73\), \(p=2\times10^{-34}\) --- the association does not depend on which, or whether any, physical predictor is included). The same 400 instances were also resubmitted on a third processor, \texttt{Advantage\_system4}, for this section's cross-processor campaign below; reanalyzing that already-collected data the same way holds at both cells
(split-embedding \(-0.53\) to \(-0.51\) at Empirical, \(-0.72\) to \(-0.64\) at Corner, every bootstrap CI excluding zero), extending the Hamming-distance replication to a second further processor beyond Zephyr, alongside \texttt{Advantage\_system6} below. That set never recorded the competing state's full spin configuration, only its Hamming distance, so the barrier
estimate could not be reanalyzed this way; for the prospective stage we separately collected 100 fresh instances per cell (seeds disjoint from every other campaign in this study; analysis plan, test and
success criteria frozen before any QPU submission) on both Zephyr and \texttt{Advantage\_system6}, saving the full configuration this time. Both
predictors hold at all four chip \(\times\) cell units on this prospective sample (partial \(\rho\) given SA occupation and each cell's preregistered predictor, \(|M|\) at Empirical and net field at Corner: Hamming, split-embedding estimate, \(-0.31\) to \(-0.72\) (same-embedding estimate \(-0.50\) to
\(-0.75\)); barrier, which has no split-embedding equivalent since it needs each embedding's own full competing-state configuration, \(-0.39\) to \(-0.67\); every bootstrap CI excludes zero). Landscape geometry's association with hardware failure is therefore not an artifact of the extreme-selected diagnostic sample it was first observed on,
though its evidentiary status remains that of an association rather than a demonstrated mechanism, in the sense made explicit below (Fig. 12). Figure 11 shows this association directly, as a partial-rank residual plot (residualizing both occupation and Hamming distance on SA occupation and, for each cell, its own preregistered predictor ---
\(|M|\) at Empirical, net field at Corner), for both cells and both evidentiary stages; it plots the same-embedding association for visual clarity (one point per instance, not one per split), so its
\(\rho\) matches the same-embedding estimates quoted above, not the more conservative split-embedding headline numbers.

\begin{figure}[tbp]
\centering
\includegraphics[scale=0.74]{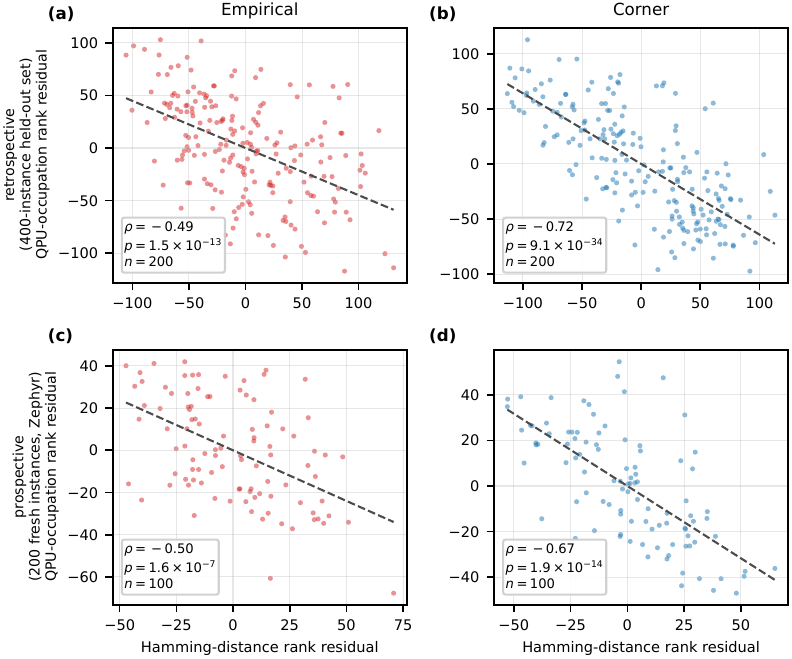}
\caption{Landscape geometry predicts an instance's own occupation beyond classical hardness and the preregistered predictor (Sec. III F), shown as rank-residualized scatter plots: both axes are ranks of
the raw quantity, residualized by regressing out SA occupation and, for each cell, its own preregistered predictor (ground-state magnetization \(|M|\) at Empirical, net field at Corner), not the raw
Hamming distance or occupation values themselves. The Hamming distance on the horizontal axis is a hardware-derived diagnostic, read from the competing state the QPU actually returned, not a
property available before running the annealer. Top row: retrospective reanalysis of the 400-instance Table II held-out set. Bottom row: the prospective, analysis-plan-fixed-before-collection
sample of 100 fresh instances per cell (Zephyr shown; SM Table S11 reports all four chip \(\times\) cell units). Dashed line, a least-squares trend in this rank-residualized space, not a claim of a
linear physical relationship between the raw quantities; \(\rho\) and \(p\) are the partial Spearman correlation and its two-sided \(p\)-value, and \(n\) is given in each panel. The association is
not driven by a few extreme instances: excluding the most distant 5\% of each panel by raw Hamming distance changes \(\rho\) only slightly (e.g.\ \(-0.49\to-0.46\) at Empirical/retrospective,
\(-0.72\to-0.72\) at Corner/retrospective, \(-0.50\to-0.42\) at Empirical/prospective, \(-0.67\to-0.64\) at Corner/prospective), with every trimmed estimate still significant at \(p<10^{-4}\). The \(\rho\)
plotted here uses the same embeddings' reads for both axes; the paper's preferred, more conservative estimate separates them (Sec. III F), giving \(-0.26\) to \(-0.35\) at Empirical/retrospective,
\(-0.61\) at Corner/retrospective, \(-0.31\) to \(-0.44\) at Empirical/prospective and \(-0.60\) to \(-0.62\) at Corner/prospective (SM Table S11).}
\end{figure}

\textbf{Whether this association is an artifact of shared reads.} The competing state and \(Y\) are not independent measurements: both are computed from the same embeddings' reads (the competing state is each embedding's modal non-ground-state read; \(Y\) is the same embedding's ground-state fraction), so a negative association between Hamming distance and \(Y\) could in principle be mechanical rather than physical --- when \(Y\) is high, the few non-ground-state reads are sparse and concentrated near the ground state (single-flip excitations), whereas when \(Y\) is low the non-ground-state pool is the bulk of the distribution and need not be geometrically close to it. We tested this on both the retrospective 400-instance set and the prospective 200-instance set (both chips), each of whose four embeddings per instance allow a clean split: Hamming distance computed from one disjoint pair of embeddings against \(Y\) computed from the other pair, in both directions, so no single read ever contributes to both quantities being correlated. The partial correlation survives everywhere, in both split directions (retrospective --- Empirical: \(-0.49\) same-embedding vs.\ \(-0.35\) and \(-0.26\) split; Corner: \(-0.72\) vs.\ \(-0.61\) and \(-0.61\); prospective, across the four chip \(\times\) cell units: same-embedding \(-0.50\) to \(-0.75\) vs.\ split \(-0.31\) to \(-0.72\); every bootstrap CI excluding zero), so the association is not purely an artifact of shared reads at either stage. It is, however, measurably inflated by them: splitting attenuates the partial \(\rho\) by roughly 15\% at Corner and 30--47\% at Empirical in the retrospective set, and by 5--38\% across the four chip \(\times\) cell units in the prospective set, with the direction and degree of this inflation itself unexplained. We therefore report the split-embedding estimates above, not the same-embedding ones, as the headline, more conservative, independently-sampled measure of this association's strength.

This test and the one above rule out one specific concern each, not the general one. The split does rule out that the association is a bookkeeping artifact of computing Hamming distance and \(Y\) from the
identical reads. It does not rule out that both quantities are shaped by one underlying, instance-specific sampling distribution over low-energy configurations: a disjoint pair of embeddings of the
same instance still samples from physically related distributions (same Hamiltonian, same chip, nearby calibration), so an instance whose landscape favors reads far from the ground state could
plausibly show that pattern in both halves of the split without the Hamming distance being an independent cause. Nothing in this study distinguishes that reading from the alternative that disjoint-read
Hamming distance carries genuine information about the realized landscape beyond classical hardness and the preregistered predictor; we are aware of no existing data in this study -- for instance, a
competing state defined a different way, such as the second-most-frequent non-ground read rather than the modal one -- that would let us test the two readings apart, and flag this as an open question
rather than one this design resolves.

\textbf{Whether near-global-flip competing states explain the result.} Because the programmed fields break the \(Z_2\) symmetry between a spin configuration and its flip, the energy cost of flipping every spin is set by field alignment, \(E(-s)-E(s)=2\sum_i h_i s_i^{\mathrm{GS}}\) --- the same combination of \(h_i\) that governs \(|M|\) and net field. Some competing states sit closer to this global flip than to the ground state itself (raw Hamming distance \(d>N/2\) for 8.6\% of embeddings at Empirical, 36.1\% at Corner, retrospective set), raising the concern that the Hamming-distance/\(Y\) association is really this flip-gap effect relabeled rather than a separate geometric signal. We tested this on the retrospective set by folding distance, \(d_{\mathrm{fold}}=\min(d,N-d)\), which treats flip-adjacent competing states as close rather than far, and by correlating the (signed) flip-gap directly with \(|M|\), net field and \(Y\). Folding barely changes the partial correlation (Empirical \(-0.49\to-0.49\) given SA and \(|M|\); Corner \(-0.72\to-0.70\) given SA and net field, this cell's own preregistered predictor; adding \(|M|\)
alongside net field at Corner changes nothing further, \(-0.70\to-0.70\)), and the folded version survives the same split-embedding test above. The flip-gap behaves exactly as the symmetry argument
predicts --- it strongly tracks each cell's own preregistered predictor and not the other (Empirical: \(\rho=-0.81\) with \(|M|\) vs.\ \(-0.09\) with net field; Corner: \(\rho=0.70\) with net field vs.\
\(-0.22\) with \(|M|\)). The negative sign at Empirical has a direct physical reading: a high-\(|M|\) ground state is one where most spins agree in sign, which tends to override the mixed-sign field
pattern and shrink \(\sum_i h_i s_i^{\mathrm{GS}}\), the flip-gap's own defining quantity. But the flip-gap only weakly correlates with \(Y\) itself (\(|\rho|\le0.17\)) and not at all with folded
distance (\(|\rho|\le0.10\), not significant at either cell). The global-flip mechanism is therefore consistent with, but does not amount to a demonstration of, \emph{why} each cell's own field
predictor works: it tracks the predictor strongly but barely touches the outcome the predictor is used for, so it is also not an alternative explanation of the geometry/\(Y\) association --- it does
not account for that association's strength either.

\textbf{Whether flip-cluster connectivity adds information beyond Hamming distance.} Hamming distance is a count, not a structural description: the same number of flipped spins could form one localized
cluster or be scattered across the graph, a distinction raised in the context of collective spin rearrangements during cyclic reverse annealing \cite{ref48}. We explored this post hoc on the
prospective sample's full competing-state configurations, the only sample in this study with that information; the association was sensitive to the statistical specification (rank-based partial
correlation versus a linear within/between-instance decomposition versus a graph-randomized null model each supported a different subset of the four chip \(\times\) cell units) and did not reach a
specification-independent conclusion. We therefore do not report it as a result here; the full exploratory analysis, including the specifications that disagreed and why, is in SM Sec. S13.

\textbf{What static errors of realistic size, as opposed to the actual measured errors, explain.} The static-error model of Sec. III D uses synthetic Gaussian offsets. To test the specific, physical
version of that account, we measured each chip's per-qubit field response in situ (uncoupled qubits, five field levels, three processors; split-half reliability \(\rho=0.67\)--\(0.83\)). The realized offsets are much
smaller than the 1\% level used as the Gaussian model's standard deviation in Secs. III D--E: the median measured \(|\text{offset}|\) is 0.05\% of the hardware \(h\)-range on Zephyr, 0.20\% on
\texttt{Advantage\_system4} and 0.30\% on \texttt{Advantage\_system6} (90th percentile 0.13--0.74\%, maximum 0.36--1.87\%). This resolves the apparent tension between the synthetic 1\% model (which
already displaces 63\% of easy instances' certified ground states) and easy instances' still-substantial occupation on hardware: the model's 1\% level is a conservative upper bound drawn from the
documented ICE range \cite{ref1}, not a measurement of the error actually realized on these qubits, which is several-fold smaller. We then asked
whether the \emph{realized} offset map, not just its typical size, predicts which embeddings and instances fail, beating 19 label-shuffled controls of the same offsets. It does not: none of 14 such
pre-registered tests, across two cells and three processors, beats its shuffled controls. The measured-offset \emph{map} is not the explanation. Offset \emph{sensitivity of the measured size},
independent of the specific pattern, remains predictive: on Zephyr and \texttt{Advantage\_system4}, fragility to offsets of realistic magnitude predicts hardware failure beyond SA about as strongly
as any shuffled version of the same offsets (partial \(\rho=0.22\)--\(0.64\)). Hardness is therefore a property of an instance's \emph{sensitivity} to realistic noise in general, not of the particular
error map realized on the chip on a given day \cite{ref2}. This sharpens, rather than weakens, the hedge already stated below: static control errors remain an association with failure, not a
demonstrated causal map.

\textbf{Table VII.} Pre-registered predictions by campaign (Sec. III C and this section). Full itemized ledger, with registration dates and confirmation rules, in SM Table S6.

\begin{center}
{\footnotesize\setlength{\tabcolsep}{4pt}\begin{tabular}{@{}lrrrl@{}}
\toprule
Campaign & Predictions & Held & Failed & What it tests \\
\midrule
\parbox[t]{0.23\linewidth}{\raggedright A1 held-out predictor test (Sec. III C)} & 4 & 2 & 2 & \parbox[t]{0.37\linewidth}{\raggedright physical predictors, 2 cells} \\
\parbox[t]{0.23\linewidth}{\raggedright Q1 day-to-day stability} & 16 & 16 & 0 & \parbox[t]{0.37\linewidth}{\raggedright ranking stability, 4 days} \\
\parbox[t]{0.23\linewidth}{\raggedright Q2 cross-processor transfer} & 12 & 9 & 3 & \parbox[t]{0.37\linewidth}{\raggedright ranking, QA\textless{}SA, predictors, 2 new chips} \\
\parbox[t]{0.23\linewidth}{\raggedright Q3 cell, graph-family and density transfer} & 10 & 9 & 1 & \parbox[t]{0.37\linewidth}{\raggedright new cells, ER graph family, density} \\
\parbox[t]{0.23\linewidth}{\raggedright Q4 gauge dependence} & 16 & 13 & 3 & \parbox[t]{0.37\linewidth}{\raggedright static bias, 2 chips} \\
\parbox[t]{0.23\linewidth}{\raggedright Q5 hard/easy diagnostics} & 8 & 7 & 1 & \parbox[t]{0.37\linewidth}{\raggedright trapping asymmetry, 2 chips} \\
\parbox[t]{0.23\linewidth}{\raggedright Q6 measured-offset map} & 17 & 3 & 14 & \parbox[t]{0.37\linewidth}{\raggedright specific vs.\ generic static-error account} \\
\parbox[t]{0.23\linewidth}{\raggedright Q7 size-scaling transfer} & 3 & 2 & 1 & \parbox[t]{0.37\linewidth}{\raggedright Fig. 2c collapse, 1 new chip} \\
\parbox[t]{0.23\linewidth}{\raggedright Q8 Corner net-field replication} & 1 & 0 & 1 & \parbox[t]{0.37\linewidth}{\raggedright chain-length-matched embeddings, design-controlled confound test} \\
\textbf{Total} & \textbf{87} & \textbf{61} & \textbf{26} & \\
\textbf{Total excl.\ Q1} & \textbf{71} & \textbf{45} & \textbf{26} & \parbox[t]{0.37\linewidth}{\raggedright near-certain-by-design tests removed} \\
\bottomrule
\end{tabular}}
\end{center}

The 87 predictions are not of comparable strength and should not be read as 87 independent odds-on bets. Q1's 16 near-certain-by-design tests (the ranking was already known to be \(\rho\ge0.989\)
stable before these were run) sit at one end; Q6's 17 tests, which ask whether a specific measured error map beats a null of the same size and fail 14 of 17, sit at the other; the remaining campaigns
(Q2--Q5, Q7) are genuine, less-certain generalization tests of the A1 result, and Q8 is a single, purpose-built replication of the weaker A1 test under a confound this study itself identified.
Removing Q1 alone brings the ledger to 45 of 71 held, against 61 of 87 with it included; we report both throughout rather
than letting the near-certain tests inflate the headline count. Only the four A1 tests of Table II are Holm-corrected as a single family; the other 83 each carry their own
pre-registered threshold, with no family-wise correction across the full ledger; most are point-threshold or compound-proportion claims (e.g.\ \(\rho\ge0.6\), \(\ge\)90\% of instances) rather than
single significance tests and are not amenable to a \(p\)-value multiplicity correction. Restricted to the 36 predictions across Q2--Q6 that are single significance tests with a reported \(p\)-value,
a Holm correction at family-wise \(\alpha=0.05\) leaves 21 of 36 significant; exactly two currently held predictions in this sub-family lose significance under the correction --- the \(|\Sigma h|\)
predictor at the Corner cell on \texttt{Advantage\_system4} (\(p=0.016\), Holm-adjusted \(0.20\)) and the ranking-transfer test at the Corner cell on \texttt{Advantage\_system6} (\(p=0.0052\),
Holm-adjusted \(0.078\)) --- while every other held prediction in this sub-family, including the reverse-annealing recovery asymmetry (Sec. III F) and the gauge-fragility association, survives with
Holm-adjusted \(p\le10^{-4}\) (full ledger and computation in SM Table S6; every held prediction's reported \(p\) is below 0.016 before this correction).

\subsection{Dynamical mechanism tests: quantum dynamics, effective temperature and time-to-solution}\label{mechanism-tests}

Four further analyses test candidate dynamical mechanisms and compare the annealer with a tuned classical solver. Each was pre-specified in its own analysis script before being run on the held-out
sample, but none is part of the formal pre-registration ledger of Table VII; full parameters, discussion and the quantitative test-by-test tables (T1--T4c; the landscape-geometry family T5--T5d of
Sec. III F is tabulated alongside them, below) are in SM Sec. S13 (SM Tables S7--S11), which also reports an exploratory, post hoc flip-cluster-connectivity test (T5e) that did not reach a
specification-independent conclusion and is not reported as a result in the main text.

A per-instance-tuned SA reaches the paper's reference success probability 25--50\(\times\) faster than the paper's own SA reference budget (T1); benchmarked against this tuned SA under two time
accountings for the annealer, the annealer is slower on every chip and cell under access-time accounting and under anneal-time-only accounting as well, except at the low-field Corner cell on
Zephyr, where pure anneal time is competitive on about one in eight instances (12\%). A short, non-equilibrated path-integral simulated quantum anneal (T2) predicts the hardware's ranking beyond
classical hardness and a classical-rotor control (T2b, partial \(\rho=0.23\)--0.41 across both tests), but this ranking signal weakens and loses significance at one of six chip\(\times\)cell units
at a ten-fold finer imaginary-time discretization and does not reproduce the hardware's absolute failure rate. A second-order transverse-field perturbation on each instance's relaxed competing
state (T3) does not discriminate hard from easy instances as a simple avoided-crossing picture predicts (Fisher \(p\approx0.98\)--0.99), ruling out that mechanism at second order. An
effective-temperature fit by parallel tempering (T4, T4b) gives a different verdict at each cell --- non-thermal at the calibrated cell (occupation-to-thermal ratio 0.35 hard vs.\ 1.57 easy), near-thermal
ground-state occupation at the low-field corner (0.57 vs.\ 1.69) --- but a held-out check on the full energy distribution's variance (T4c) finds neither cell well described as a single-temperature
Gibbs state, so ``close to thermal'' at Corner is agreement in the first moment only, not in the sampled distribution's shape.

\textbf{Table VIII.} Mechanism-test verdicts at a glance (full tables, SM Tables S7--S11).

\begin{center}
{\scriptsize\setlength{\tabcolsep}{6pt}\begin{tabular}{@{}lll@{}}
\toprule
Test & Mechanism tested & Verdict \\
\midrule
T1 & \parbox[t]{0.58\linewidth}{\raggedright Annealer beats a per-instance-tuned classical solver (time-to-solution)} & \parbox[t]{0.26\linewidth}{\raggedright Not supported, except pure anneal time at Corner/Zephyr} \\
T2, T2b & \parbox[t]{0.58\linewidth}{\raggedright Non-equilibrium quantum dynamics ranks instances beyond classical hardness} & \parbox[t]{0.26\linewidth}{\raggedright Partially supported: ranking signal present, sweep-count sensitive} \\
T3 & \parbox[t]{0.58\linewidth}{\raggedright Simple avoided-crossing picture favors the trap over the ground state} & \parbox[t]{0.26\linewidth}{\raggedright Not supported} \\
T4, T4b, T4c & \parbox[t]{0.58\linewidth}{\raggedright A single effective temperature describes hardware sampling} & \parbox[t]{0.26\linewidth}{\raggedright Partially supported: cell-dependent in the mean, non-thermal in the full distribution at both cells} \\
T5--T5d & \parbox[t]{0.58\linewidth}{\raggedright Competing-state geometry (Hamming distance, barrier) predicts occupation beyond classical hardness and the preregistered predictor (Sec. III F)} & \parbox[t]{0.26\linewidth}{\raggedright Supported as an association; not yet a demonstrated mechanism} \\
\bottomrule
\end{tabular}}
\end{center}

\section{Discussion}\label{discussion}

The principal finding is a reproducible association between hardware-observed competing-state geometry and ground-state occupation that persists after conditioning on classical hardness and each
cell's preregistered physical predictor, and survives both independently-sampled (split-embedding) and prospectively-collected validation (Sec. III F). This association does not establish a causal
mechanism, and it is distinct in kind from the weaker, regime-dependent pre-registered predictors discussed next.

The results also support three further observations. First, at fixed ensemble parameters, the annealer's ground-state occupation varies across nearly its full range between instances (Sec. III A). This instance
heterogeneity is consistent with, though distinct from a direct demonstration of, the J-chaos signature documented in the spin-glass benchmarking literature and on native hardware instances by Zhu \emph{et al.} \cite{ref5} and Pearson \emph{et al.} \cite{ref2} -- J-chaos specifically denotes sensitivity of the ground state to small coupling perturbations, which we do not test directly here; what we observe is the resulting wide spread in occupation across instances, on structured, embedded instances rather than hardware-native synthetic ones. Second, part of this variation is predictable before running
the hardware. A physical feature of the exact ground state predicts failure on held-out data, and a static-error model adds information beyond classical hardness at the low-field corner, though not
at the calibrated cell (Secs. III C--D). Third, the predictive
feature is not universal, and the surviving one does not transfer: magnetization works at the calibrated cell and is confirmed there, but not at the other cell; net field appeared to work at the
low-field corner under the original pre-registration but failed a larger, dedicated, confound-controlled replication (Sec. III C).

Three phenomena co-occur in the hard instances: reproducible ground-state failure that spans nearly the full occupation range (Sec. III A), sensitivity to perturbations at the hardware's documented
error scale (Secs. III C--E), and a schedule-dependent dynamical trapping signature under reverse annealing, now measured on 48 instances and a second processor (Secs. III E--F). The data below show
these three co-occurring in the same instances and, for the trapping signature, reproducing across chips; they do not show that the three share one underlying cause, and the paragraphs that follow
treat each as a candidate contributor rather than as an established mechanism.

Figure 12 collects these relationships and marks each by evidentiary status, since the paper's findings are not uniform in kind. \emph{Established findings} (solid black) are associations that
reproduce everywhere tested and do not depend on any proposed mechanism, though not all were tested on the same axes: classical hardness predicts QA occupation across cells, chips and graph
families (Sec. III F), and landscape geometry (Hamming distance and, Sec. III F, an estimated energy barrier) corroborates the trapping signature and predicts the instance's own occupation beyond
classical hardness and the preregistered predictor, first on the Q5 diagnostic subsample (partial \(\rho=-0.59\)) and, confirming that this was not an artifact of its extreme-selected sampling, again on
samples never selected on \(Y\) at every chip and cell tested (Sec. III F) --- geometry has not been tested on the new graph family or density of Sec. III F, so this reproducibility claim is scoped
to cells and chips only. \emph{Statistical predictors} (blue) are the two pre-registered physical features, only one of which retained support in subsequent testing (Sec. III C); they are real, held-out
associations, each confirmed at one cell only, and modest in absolute terms (Sec. III C's triage-policy check finds little benefit over a random baseline of the same size at this effect size).
\emph{Proposed mechanisms} (red, dashed or dotted) are candidate explanations that are consistent with the data but not demonstrated as causes: the static-error account offers a proposed interpretation of why the two
predictors predict -- a weaker claim for net field, given its failed replication, than for magnetization -- without identifying the realized error map (Sec. III D, III F), and dynamical trapping co-occurs with hardness without being shown to be its cause (Sec. III E--F) --- geometry
predicting occupation is a stronger association than mere co-occurrence, confirmed beyond its original small, extreme-selected sample, and is still not evidence that geometry \emph{causes} the failure rather than
sharing a common unobserved origin with it. Keeping these three categories visually and verbally distinct is the organizing principle of the sections that follow.

\begin{figure*}[tbp]
\centering
\includegraphics[max width=\linewidth]{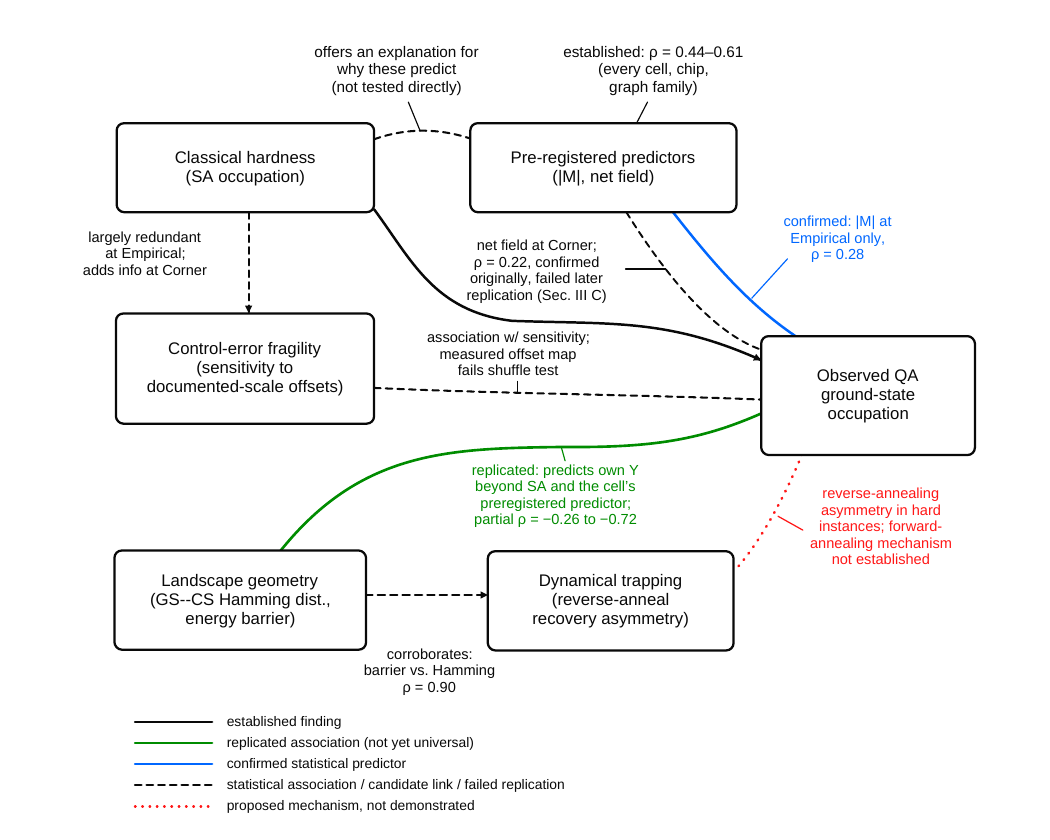}
\caption{Synthesis: how the paper's instance properties, candidate mechanisms and the observed hardware outcome relate, marked by evidentiary status. Solid black, an established finding that
reproduces everywhere tested (classical hardness, across cells, chips and graph families). Solid green, a replicated association that is not yet established to the same degree (landscape geometry:
confirmed across cells, chips and both retrospective and prospective samples, but inflated by shared reads and not yet tested on a new graph family or density). Solid blue, a confirmed,
pre-registered statistical predictor that survives (\(|M|\) at Empirical only). Dashed gray, a statistical association or candidate explanatory link not demonstrated to be causal, or a once-confirmed
predictor that failed a later replication (net field at Corner, pre-registered and confirmed originally but not supported by a larger, dedicated, confound-controlled follow-up; Sec. III C). Dotted red,
a proposed mechanism that co-occurs with hardness but has not been shown to cause it.}
\end{figure*}

Under the social-network reading of the model noted in Sec. I (local fields as individual predispositions, antiferromagnetic bonds as antagonistic ties), Fig. 4's pattern suggests network
structure plays a conditional role: when predisposition dominates over social pressure (\(r_h\) high), the fraction of antagonistic ties has little bearing on whether the system settles into a
globally coherent configuration, whereas when social pressure dominates (\(r_h\) low) that fraction, and how densely connected individuals are, both matter sharply. This is an interpretation of a
hardware-sampling result, not a claim this study tests directly against opinion data.

The diagnostics connect to known mechanisms, though not in the simplest way. The hard instances' competing states lie a median 10 spins from the ground state, against 1 spin for the easy instances.
That is the landscape signature of hardness identified by Martín-Mayor and Hen \cite{ref4}, the same large-versus-small Hamming distance separating hard from easy instances' excited states first
reported on the earliest hundred-qubit devices \cite{ref3,ref42} --- our result reproduces that signature on embedded, structured instances with exactly certified ground states, rather than
discovering it, and sharpens it with an exploratory, per-instance conditional association subsequently confirmed in an independently collected, prospectively specified validation experiment (Sec. III F) that the original comparison did not attempt. It is also consistent with the broader statistical-mechanics finding that competing cooperative and frustrating interactions reorganize an Ising system's ground-state landscape \cite{ref35}, seen here in instances that are easy for SA. The hard instances' ground state is stable under reverse annealing at
\(s\ge0.35\) but lost at \(s\le0.30\), and is not regained from the competing state. That pattern is consistent with a mid-anneal avoided crossing between the two states, the small-gap mechanism of
Refs. \cite{ref3,ref24,ref45}, perturbative crossings late in the anneal being one proposed source of exactly this kind of heavy-tailed hardness on degenerate instances \cite{ref49}, and with
freeze-out of the dynamics before the end of the anneal \cite{ref7}. A direct test of the simplest avoided-crossing picture argues against it (T3, Sec. III G): a
second-order transverse-field perturbation around each instance's relaxed competing state does not favor the hardware's observed trap over the ground state more often for hard instances than for easy
ones, and the trap is neither the lowest-energy nor the most quantum-favored state among its near-degenerate neighbors. This rules out a second-order perturbative avoided-crossing account, evaluated at
each instance's own relaxed hardware-observed competing state, on the Q5 (24+24, two chips) and seed-42/A2 (6+6) samples tested here; it does not rule out freeze-out,
a higher-order version \cite{ref43}, or an instance-dependent effective temperature \cite{ref25} (see below). Degenerate ground states, which transverse-field annealers sample unfairly \cite{ref26,ref37,ref38}, are rare here. Among the configurations sampled here, the ground state is unique for 186 of the 200
instances at the Empirical cell and for 11 of the 12 in Table V; this is uniqueness among sampled configurations, not a certified statement that the ground state is non-degenerate (Sec. II C records 14 instances
at Empirical and 8 at Corner with more than one certified ground-state configuration).

A static-error account offers a proposed interpretation, not an established explanation, of why the two pre-registered features are predictive -- a more tentative claim for net field, which did not
survive its dedicated replication (Sec. III C), than for magnetization. If programmed fields carry fixed offsets, the ground state should be most fragile when its stability
rests on fields that are small or nearly cancel. At the low-field corner, where couplings dominate, the net field sets how strongly the ground state is selected among nearly degenerate low-energy
states. At the calibrated point, where fields are larger, magnetization indicates how much the ground state relies on field alignment. This interpretation is consistent with the data. Static offsets
of ICE size do make the hard instances' ground states more fragile (Sec. III E), but they do not by themselves single out the competing states the hardware finds. We measured the realized on-chip
\(h\) offsets for this purpose, following the method of Zaborniak and de Sousa \cite{ref27} (Sec. III F): on three processors, the \emph{specific} measured offset map does not predict which embeddings
or instances fail beyond label-shuffled controls of the same offsets, while offset \emph{sensitivity of the measured size}, independent of the specific pattern, does. The static-error account
therefore explains why certain instances are fragile to perturbations of documented magnitude in general, not which particular error map the hardware happens to realize on a given day.

The evidence for static control errors is accordingly an association with offset sensitivity, not a causal identification of the realized error map (Sec. III F). A perturbation model that ranks
instances by their sensitivity to \emph{any} small change would correlate with hardware failure even if the hardware's failure came from another source. Freeze-out of the transverse-field dynamics
before equilibration \cite{ref7}, an avoided crossing \cite{ref24}, or chain-break-induced decoherence of logical spins could act alongside ICE. Weaker chains helping, as found here, matches the
instance-dependent optimal chain strength reported by Gilbert and Louise \cite{ref28}. The hardware diagnostics narrow, but do not eliminate, these alternatives. Anneal time and pauses barely move
occupation, which argues against a purely kinetic explanation at these time scales. Stronger chains hurt rather than help. The competing state recurs across embeddings and survives most gauges.

Two further mechanism tests (T2--T2b, T4--T4b; Sec. III G) bear on the dynamical picture. A short, non-equilibrated simulation of the quantum dynamics predicts the hardware's ranking beyond
classical hardness and beyond a classical-rotor-model control \cite{ref33} — evidence for a ranking component that neither a classical sampler nor a semiclassical rotor model captures — though it
does not reproduce the hardware's absolute failure rate and is not proof of coherent quantum dynamics, since the simulation itself is a classical approximation to the quantum evolution; the
ranking signal itself weakens, and loses significance in one of six chip-cell units, at a ten-fold longer imaginary-time discretization of the same simulation (Sec. III G), so it should be read as
sensitive to simulation detail rather than as a robust quantum signature. Separately,
fitting each instance's effective temperature by parallel tempering gives a thermal-equilibrium account whose \emph{ground-state occupation} fits well at the low-field Corner cell but leaves a clear
non-thermal excess at the calibrated Empirical cell, where hard instances are under-sampled and easy instances over-sampled relative to the fitted thermal state. This agreement is in one moment only:
a held-out check on the energy variance (Sec. III G) finds the hardware's full energy distribution is markedly more spread out than the fitted single temperature predicts at \emph{both} cells, so
Corner is better described as thermal in its ground-state occupation than as a noisy Gibbs sampler at one well-defined temperature \cite{ref47}, with extra, non-thermal trapping concentrated on the hard instances at
Empirical. Taken together, the hardware's instance ranking carries a component that looks like a non-equilibrated quantum anneal, a component whose ground-state occupation (but not its full energy
distribution) resembles thermal sampling at an instance-specific temperature (strongest at Corner), and a residual, non-thermal loss concentrated on the hard instances at Empirical; the three phenomena
named at the start of this section are better read as converging symptoms of several only partly overlapping mechanisms than as the signature of one.

If a single most plausible picture is to be drawn from these partly overlapping results, it is this: hardness in this ensemble tracks the \emph{margin} by which the true ground state is favored over
its most competitive alternative, rather than any one noise channel or dynamical process. An instance whose ground state wins only narrowly -- by a small net field or a magnetization close to the
competing state's, in the sense already used for the static-error account above -- is, on this reading, the same kind of instance whose landscape geometry shows a short Hamming distance or a low
energy barrier to that alternative (Sec. III F), and the same kind whose occupation is most sensitive to perturbations of realistic size regardless of the specific offset map realized (Q6, Sec. III F).
A narrow margin is, in this view, not itself a mechanism but a shared property that several different proximate mechanisms could act on: it would make an avoided crossing narrower and therefore more
consequential if it occurs, make freeze-out more likely to leave the system on the wrong side of it, and make small static or thermal perturbations more likely to favor the competitor. This reading is
consistent with every result above without requiring any of the individual mechanism tests to have succeeded on its own, and it is why no single test here -- T2/T2b, T3, T4/T4b, or the measured-offset
map -- isolates one dynamical cause: each probes a different way a narrow margin could fail to be realized, not a different, competing hypothesis for what the margin is. It remains an interpretation of
the pattern across tests, not a finding any one of them establishes, and a direct test would need to manipulate the margin itself, independent of the features used to estimate it here, which this study
does not attempt.

This ensemble-level, per-instance approach is complementary to single-qubit studies of the hardware's intrinsic noise. Morrell \emph{et al.} \cite{ref32} isolate a single qubit to discriminate
thermal and longitudinal-field noise models from the annealer's own dynamics. The present data cannot adjudicate between those mechanisms at that level; what they add is that, whatever the
qubit-level noise processes are, their net effect on a many-spin embedded problem is not uniform across instances but tracks static, pre-registered features of the problem itself.

Three methodological implications follow. First, annealer-versus-classical comparisons drawn from one or a few instances are unreliable in this regime. This study's own single-instance and six-seed
readings showed annealer advantages that larger samples reversed (Table I; SM S2). Second, the comparison depends on how SA itself is budgeted. The paper's SA reference (5,000 reads, 1,500 sweeps) is
a high-quality \emph{sampler} baseline, not a tuned \emph{solver} baseline: a per-instance-tuned SA reaches the same success probability 25--50\(\times\) faster (T1, Sec. III G). Benchmarked
against this tuned SA, the annealer is slower under every time accounting tested, including pure anneal time alone except at the low-field Corner cell on Zephyr. ``The annealer falls behind SA'' should therefore be read as an occupation statement against a fixed, generously budgeted sampler; against a tuned solver reference, on wall-clock
time, the disadvantage is if anything larger. Third, the scaling result should be read narrowly: at default settings and a fixed anneal, the annealer's ground-state sampling probability falls faster
with \(N\) than that of a fixed-budget SA. This is not a time-to-solution measurement in itself, and neither this nor the tuned-SA comparison above establishes a general quantum-versus-classical
separation; both concern sampling a fixed, un-optimized Hamiltonian under one schedule. Both comparisons are specific to the classical algorithm used (simulated annealing, not a tuned exact or
specialized heuristic solver), its tuning procedure (per-instance sweep-count tuning against the \(\mathrm{TTS}_{99}\) criterion), the success criterion itself (99\% ground-state sampling probability),
and the hardware time accounting applied (anneal time alone versus full access time); a different classical competitor, tuning budget, success threshold or accounting convention could shift the
comparison, and nothing here should be read as a claim about quantum computational advantage in general.

A practical implication would follow from the predictability itself, independent of mechanism, if either preregistered predictor could serve as a cheap, pre-hardware screen --- but neither can, for
different reasons. Magnetization, the stronger predictor at the calibrated cell, is computed from the exact ground state itself, so it cannot serve this role for an unsolved problem: by the time the
ground state is known, the pre-screening question is moot. Its practical use is as a retrospective diagnostic instead, for example to decide which already-solved instances are worth re-submitting to
hardware or are safe to report as benchmarks, not as a pre-screen for new, unsolved ones. Net field is computed directly from the programmed fields, with no optimization required, so it would not
have this particular limitation --- but it is also the predictor that failed the larger, confound-controlled replication above (Sec. III C), so it offers no pre-screening value either. The same
distinction applies to comparing hardware settings by either feature: that is a benchmarking use, on instances with known answers, not a production one.
Stated plainly: the predictability demonstrated here is real but currently too weak, and too regime-dependent, to support operational pre-screening of any kind; the study's more robust practical lessons are
methodological rather than predictive — do not benchmark an annealer against a few instances, certify references exactly rather than trusting a heuristic's best, and budget any classical comparator
explicitly rather than at one fixed, arbitrary setting.

These findings carry several limitations of scope. The main results (Figs. 1--3, 5--7, Tables I--VI) come from one processor (\texttt{Advantage2\_system1}, Zephyr); Sec. III F cross-checks the central
claims on two further processors, but not every analysis -- Q6's measured-offset test and the Q4 gauge campaign cover three chips, the hard/easy diagnostics and the size-scaling collapse cover two,
and the static-error model of Sec. III D was not re-fit per processor. Data were collected across a period of live access spanning possible recalibrations, which were not tracked within a run and
could contribute to instance-to-instance variability; Sec. III F's day-to-day campaign bounds this directly (ranking stable at \(\rho\ge0.989\) across four days). The pre-registered predictor test
of Sec. III C covers two cells and two predictors; Sec. III F extends it to three further cells and a new graph family, but other candidate predictors (classical gap, basin count, realized
antiferromagnetic fraction) examined only in the exploratory phase were not tested at scale in either campaign. Two cells at \(\alpha=0.05\) (\(m=2\), \(m=5\), \(r_h=0.25\)) have only six-instance
comparisons (Table I, †). SA was run at one fixed budget (5,000 reads, 1,500 sweeps); its own occupation shifts with tuning (mean 4.4 pp across four budgets spanning 40\(\times\) in effort), without
changing the sign of any comparison (SM S9). Most annealer conditions used 1,000 reads per embedding; at 2,000 µs, 300 reads were used in the sweep (SM S11) and 400 in the Sec. III E diagnostics, to
respect the per-problem access-time limit.

The mechanistic account has corresponding limits. The static-error model of Sec. III D is classical and uses Gaussian offsets of documented magnitude, not the measured on-chip errors for these
instances. Sec. III F's direct measurement of the realized offsets (Q6) found that the specific, physical offset map does not predict failure beyond shuffled controls of the same offsets on any of
three processors; the static-error account that survives is therefore about sensitivity to offsets of a given size, not a validated physical mechanism. The chip-quadrant test (Sec. III E) excludes
region-specific crosstalk as the cause of failure for the 12 instances tested, but not crosstalk effects that are uniform across the chip. The hard/easy comparison of Sec. III E (Table V) uses six
instances per group, selected by a rule fixed in advance but from the extremes of the held-out distribution; the Sec. III F replication (24 + 24 fresh instances, two processors) is four times larger
but uses the same selection rule, so both characterize the mechanism at the extremes of the held-out distribution, not its prevalence across the full ensemble.

Several statistical caveats also apply. The 87 predictions of Table VII were each pre-registered with their own threshold, not Holm-corrected as a single family (only the four A1 tests of Table II
were); the family-wise Holm correction restricted to the 36 single-significance-test predictions (21 of 36 survive) is reported next to Table VII in Results, not repeated here. The net-field predictor at Corner, the weaker of the two confirmed A1 tests, loses significance at the pre-registered confirmation threshold once chain length and chain-break fraction are
controlled for, and failed to replicate outright in a larger, purpose-built, confound-controlled follow-up (Sec. III C); magnetization at Empirical is essentially unaffected by the same conditioning. The realized
autoscaling factor (chain-strength divisor) makes the weakest programmed bonds comparatively imprecise under ICE, but both predictors survive conditioning on chain strength and the weak-bond
fraction, so this specific confound does not appear to explain either association. T1--T4c (Sec. III G) and T5--T5d (Sec. III F) were pre-specified in their own analysis scripts before being run on
the held-out sample, but are not part of the formal, time-stamped pre-registration ledger of Table VII and SM Table S6, and should be read with that weaker evidentiary status in mind. A further, fully
post hoc exploration of flip-cluster connectivity (SM Sec. S13) did not reach a conclusion stable to the choice of statistical specification and is reported only in the Supplement, not as a result
here. The SQA test
(T2) used the simulator library's default transverse-field schedule and Trotter-slice count rather than an independently chosen one; the schedule itself was not varied, and at a ten-fold longer
imaginary-time discretization the ranking signal weakens and loses significance at one of six chip-cell units (Corner, \texttt{Advantage\_system6}), so the headline 6/6 result is specific to the
100-sweep setting tested. The classical-rotor control's own signal (T2b) is sensitive to the rotor temperature, holding at the primary temperature but not at both lower temperatures tested. Finally,
at the calibrated Empirical cell, a cheap-feature machine-learning baseline (Sec. III C) matches \(|M|\)'s held-out discrimination, so \(|M|\) is not a meaningful advance over a simple
antiferromagnetic-fraction model there; the only cell where a physical predictor beat cheap features at all was the low-field Corner cell, and that advantage belonged to net field --- the same predictor
that failed the dedicated replication above. No pre-registered physical predictor, at either cell, demonstrates a replicated advantage over a cheap, hardware-agnostic frustration proxy.

\section{Methods}\label{methods}

\textbf{Software.} \texttt{dwave-system} 1.33.0 (\texttt{DWaveSampler}, \texttt{FixedEmbeddingComposite}), \texttt{minorminer} 0.2.19 (\texttt{timeout=90}, \texttt{tries=10}), \texttt{dimod} 0.12.21,
\texttt{dwave-neal} 0.6.0 (SA), \texttt{networkx} 3.2.1, \texttt{numpy} 2.0.2, \texttt{scipy} (statistics), Gurobi 12.0.3 (exact ground states for \(N\le180\); \texttt{MIPGap=0}, \texttt{MIPGapAbs}
\(=10^{-9}\)) and OR-Tools CP-SAT 9.15 (exact ground states for \(N=360\)). Embedding seeds, disorder seeds and SA seeds are listed in the Supplemental Material. The
SA sampler is seeded with the disorder seed. It draws from its own generator, independent of the one that builds the instance.

\textbf{Processors and data-collection period.} Three D-Wave cloud solvers were used: \texttt{Advantage2\_system1} (Zephyr; main results and A1), \texttt{Advantage\_system4} and
\texttt{Advantage\_system6} (Pegasus; Sec. III F only). Each submission recorded its solver's live hardware-graph snapshot identifier (logged alongside the raw results and available with the data,
per Data and code availability below); because D-Wave cloud solvers are recalibrated periodically and do not expose a single static version number, this per-submission snapshot identifier, not a
fixed version string, is the correct unit of reproducibility here and is the same information Sec. III F's day-to-day campaign uses to bound recalibration drift. Hardware data collection ran from
2026-10-08 (first A1 submissions) to 2026-11-01 (final Q7 submissions, Sec. III F), with one later addition: the prospective landscape-geometry validation's 200 fresh instances (Sec. III F) were
submitted on 2026-11-12, after this window, on \texttt{Advantage2\_system1} and \texttt{Advantage\_system6}.

\textbf{Exact certification.} Each instance is rebuilt from its seed, and Eq. (1) is solved as a binary quadratic program (Gurobi for \(N\le180\), CP-SAT for \(N=360\)). The solver status, the certified optimum and its difference from the SA reference are recorded for the pre-registered sample, the parameter sweeps and the six-seed campaign. The sweep analysis applies the certified energies to the 27 instances whose SA reference is not the ground state. The partial correlations of Table III and the static-error test of Sec. III E use the same certified references.

\textbf{Statistics.} Occupation is weighted by \texttt{num\_occurrences}. Per-instance annealer occupation is the mean over embeddings. Paired annealer-versus-SA comparisons use Wilcoxon signed-rank
tests, Holm-corrected within each sweep, with bootstrap 95\% confidence intervals on the mean paired difference. Each sweep also has one pre-specified axis-level endpoint test (paired, or
Mann--Whitney for \(N\), whose seeds are not shared across sizes). Pre-registered tests are one-sided Spearman correlations with Holm correction across the four tests. Confidence intervals on \(\rho\)
are bootstrap intervals; throughout the paper, bootstrap confidence intervals are two-sided even where the underlying hypothesis test is one-sided.

\textbf{Mechanism tests (Sec. III G).} The second-order transverse-field test (T3) relaxes each instance's hardware-observed competing state by greedy descent to its local minimum, then computes the
perturbative energy shift \(E(s;\Gamma)\approx E_\text{cl}(s)-\Gamma^2\sum_i 1/\Delta E_i\) for the Q5 (24+24, two processors) and A2 (6+6) competing states, and asks whether the shift favors the trap
over the ground state. The path-integral SQA test (T2) uses D-Wave Ocean's \texttt{PathIntegralAnnealingSampler} on the logical (unembedded) problem, 200 reads at 100 and 1,000 sweeps, on all 400
held-out instances at Corner and Empirical, at the library's default transverse-field schedule and Trotter-slice count, which were not independently varied here. The classical-rotor control (T2b, SSSV) represents
each spin as a rotor \(\theta_i\in[0,\pi]\) with energy \(E(\theta;s)=-A(s)\sum_i\sin\theta_i+B(s)[\sum_i h_i\cos\theta_i+\sum_{ij}J_{ij}\cos\theta_i\cos\theta_j]\), where \(h_i\), \(J_{ij}\) here are the
solver-convention programmed fields and couplings (sign convention \(h'=-h\) relative to this paper's own \(h\), \(J\) as returned by the instance generator and used everywhere else), linear schedule \(A=1-s\),
\(B=s\) \cite{ref33}, 1,000 Metropolis sweeps, 200 reads, at three temperatures (\(T=0.05,0.1,0.2\)); the readout is \(s_i=\mathrm{sign}(\cos\theta_i)\). \(T=0.2\), the value whose mean occupation is
closest to the hardware's, is treated as primary; \(T=0.05,0.1\) are reported as a robustness check. The parallel-tempering effective-temperature fit (T4b) uses 32 replicas with \(\beta\) on a
geometric ladder from 0.1 to 50, 2,000 burn-in and 4,000 measured sweeps per replica, on all 400 held-out instances; an earlier fixed-\(\beta\) Metropolis fit (T4) is superseded because its \(\beta\)
estimate saturated at the search bracket's cap for roughly half the instances. The tuned-SA time-to-solution benchmark (T1) defines \(\mathrm{TTS}_{99}=t_\text{read}\cdot\ln(0.01)/\ln(1-p)\) for
success probability \(p\) at sweep count \(s\in\{10,30,100,300,1000,1500\}\) (1,000 reads each), and compares the per-instance optimum against the annealer under two time accountings: anneal time
alone (20 µs/read) and full access time (programming and readout included, \(\sim\)150--190 µs/read, read from the solver's own timing fields).

\textbf{Landscape-geometry validation (Sec. III F).} The claim that competing-state Hamming distance and an energy-barrier estimate predict an instance's own occupation beyond classical hardness and the
preregistered physical predictor was tested in two stages with different evidentiary status. The \emph{retrospective} stage reanalyzes the 400 Table II held-out instances, collected for the pre-registered
predictor test of Sec. III C and reused here without new data collection; it covers Hamming distance only, since that set never recorded the competing state's full spin configuration. The
\emph{prospective} stage collects 100 fresh instances per cell (seeds 5000--5099, disjoint from every other campaign in this study) on both Zephyr and \texttt{Advantage\_system6}, saving the full
competing-state configuration needed for the barrier estimate. These 200 instances were exactly certified by the same Gurobi procedure as every other instance in this study (Sec. II C); all 200 matched
the SA reference exactly. The test (partial Spearman \(\rho\) of Hamming distance and of barrier against \(Y\), controlling for SA occupation and each cell's preregistered predictor --- \(|M|\) at Empirical, net
field at Corner --- with a bootstrap-CI-excludes-zero success criterion at each of the four chip \(\times\) cell units) was frozen before any QPU submission in this stage. The energy-barrier estimate itself is a zero-temperature,
least-bottleneck flip-path search (40 restarts, half greedy/steepest-descent and half random-order as an upper-bound control) between the ground state and the competing state; it is an upper bound on
the true minimal-energy barrier, not an exact one, since an exhaustive search is intractable at this system size. A further, fully post hoc exploration of flip-cluster connectivity on the same sample
is described in SM Sec. S13; it is not reported as a result in the main text (see Discussion).

\textbf{Data and code availability.} All scripts, raw hardware results, exact ground-state certificates and figure data are available from the author upon request. This includes the hardware graph snapshot used for embedding and the pre-registration record for every held-out test in Table II and Table VII, each held in version control with a commit timestamp that predates the corresponding data collection. A versioned, DOI-assigned public archive of the same materials, including the full pre-registration commit history, will accompany journal submission, so that every registration date and confirmation rule in Table II, Table VII and SM Table S6 is independently verifiable rather than attested only by the author.

\end{document}